\documentclass[10pt]{article}
\usepackage{graphicx,epstopdf}
\usepackage{amsmath}
\usepackage{amssymb}
\usepackage{subfig}
\usepackage{multirow}
\usepackage[top=2cm, bottom=2cm, outer=2cm, inner=2cm]{geometry}
\providecommand{\keywords}[1]{\textit{Keywords---} #1}
\def\pslash{p\!\!\!\slash }
\def\qslash{q\!\!\!\slash }
\def\xslash{x\!\!\!\slash }
\begin{document}

\title{Tensor form factors of the octet hyperons in QCD}

\author{
A. Kucukarslan$^1$ \thanks{ e-mail: akucukarslan@comu.edu.tr},
 U. Ozdem$^1$ \thanks {e-mail: ulasozdemm@gmail.com (Corresponding author)},
A. Ozpineci$^2$  \thanks {e-mail: ozpineci@metu.edu.tr} \\
\\
  \small$^1$ Physics Department, Canakkale  Onsekiz Mart University, 17100 Canakkale, Turkey\\
 \small $^2$ Physics Department, Middle East Technical University, 06531 Ankara, Turkey}

 \date{}

\maketitle
\begin{abstract}
Light-cone QCD sum rules to leading order in QCD are used to 
investigate the tensor form factors of the $\Sigma-\Sigma$, $\Xi-\Xi$ 
and $\Sigma-\Lambda $ transitions in the range 
$1~GeV^2\leq Q^2 \leq 10$ $GeV^2$.
The DAs of $\Sigma$, $\Xi$ and $\Lambda$ baryon have been 
calculated without higher-order terms. Then, studies including 
higher-order corrections have been done for the $\Sigma$ and $\Lambda$ baryon. 
The resulting form factors are obtained using these two DAs.
We make a comparison with the predictions of the chiral quark soliton model.
\end{abstract}
\keywords{Tensor form factors, Octet-octet hyperon transitions, Light cone QCD sum rules}

\section{Introduction}
\hspace*{0.5cm}How the hadrons are built from quarks and gluons, the fundamental degrees  of freedom of QCD, is one of the main open questions 
in the theory of strong interactions. An efficient way to probe the hadron structure is to study 
the hadron form factors as these quantities include direct information about the hadron structure. 
Therefore, the form factors have recently received considerable attention both in theory and experiment.
Like other form factors, the tensor form factors encode the important information about the quark-gluon structure of baryons. 

The quark distributions in the leading twist are given by the unpolarized 
distribution $f_1(x)$,
helicty distribution $g_1(x)$ and transversity distribution $h_1(x)$ function 
of the quark.
One of these functions, the transversity distribution -which describes the probability
of finding
a transversely polarized quark with longitudinal momentum fraction $x$ in
a transversely polarized baryons~\cite{Jaffe:1991kp} by its chiral-odd nature- 
is not easy to measure. The strong interactions have approximate chiral 
symmetry and electroweak interactions conserve the chirality, 
so the transversity distributions cannot be extracted in inclusive deep 
inelastic scattering (DIS). It needs to couple to another 
chiral-odd quantity in the cross-section. It can be obtained Drell-Yan processes 
and semi-inclusive deep inelastic scattering (SIDIS), because distributions 
of transversity
 do come out at leading twist in the cross-section.
In Ref.\cite{Anselmino:2007fs}, it was extracted using the experimental 
data on azimuthal asymmetries in SIDIS, from BELLE \cite{Abe:2005zx}
as well as data for the nucleon from the HERMES \cite{Airapetian:2004tw} and 
COMPASS \cite{Ageev:2006da} collaborations. Subsequently, in Ref.\cite{Cloet:2007em}, 
isovector nucleon tensor charge obtained $H_T(0) = 0.65^{+0.30}_{-0.23}$ at a 
renormalization scale $\mu^2 = 0.4~GeV^2$.
In Ref.\cite{Anselmino:2008jk} Anselmino et al. updated their nucleon isovector 
tensor charge result $H_T(0) = 0.77^{+0.18}_{-0.38}$ at renormalization 
scale of $\mu^2 = 0.8~GeV^2$. On the theoretical side, isovector tensor 
charge of the nucleon has been studied in the framework 
of lattice QCD \cite{Hagler:2007xi, Gockeler:2006zu, Bhattacharya:2015wna}, 
the chiral quark soliton model $(\chi$QSM) \cite{Lorce:2007fa, Ledwig:2010tu},
quark model \cite{Schmidt:1997vm, pasquini:2005dk}, Skyrme model \cite{Olness:1992zb},
axial vector dominance model \cite{Gamberg:2001qc} and 
dihadron production \cite{Pisano:2015wnq}
QCD sum rules and light cone 
QCD sum rules \cite{He:1994gz, He:1996wy, Aliev:2011ku, Erkol:2011iw}. 
For octet hyperon isovector tensor transition form factors 
have been studied chiral quark soliton model $(\chi$QSM) \cite{Ledwig:2010tu}.

The hyperon sector is interesting because it provides for an ideal system in which to study
$SU(3)$  flavor symmetry breaking by replacing up or down quarks in nucleons with strange ones \cite{Lin:2008rb}.
Tensor form factors play an important role in our understanding of the “tomography” of baryons. 
The hyperon tensor form factors a missing part of the this tomography. 
In the present work, we calculate the isovector tensor transition form factors of the $\Xi-\Xi$, $\Sigma-\Sigma$ and $\Sigma-\Lambda$.
In order to study the tensor form factors, one needs to use a nonperturbative method.
One of the most powerful nonperturbative methods is traditional QCD sum rules (QCDSR), which is more reliable and predictive in calculating the
properties of hadrons \cite{Shifman:1978bx, Shifman:1978by, Reinders:1984sr, Ioffe:1983ju}.
An alternative to the traditional QCD sum rules is the light cone QCD sum rules (LCSR)~\cite{Braun:1988qv, Balitsky:1989ry, Chernyak:1990ag}.
In this method, the hadronic properties are expressed with regards to the properties of
the vacuum and the light cone distribution amplitudes of the hadrons in the process. Since the form factors are
expressed with regards to the properties of the QCD vacuum and the distribution amplitudes, any uncertainty in these
parameters reflects to the uncertainty of the predictions of the form factors.

Our aim in this work is to study the tensor form factors in the framework of light-cone QCD sum rules for the octet hyperon.
We give the formulation of the hyperon form factors on the light cone and derive our sum rules. In the last section,
we present our numerical analysis and discussion.

 \section{Formalism}

 The matrix element of the tensor current between two hyperon states is parametrized by three form factors as presented  \cite{Hagler:2009ni, Gockeler:2006zu}

  \begin{eqnarray}\label{mat}
\langle H(p^\prime)|j_{\mu\nu}|H(p)\rangle&=&
 \bar{u}(p^\prime)\left[i\sigma_{\mu\nu} H_T(q^2) +\frac{\gamma_\mu q_\nu-\gamma_\nu q_\mu}{2 m_H} E_T(q^2)
 +\frac{\tilde P_\mu q_\nu-\tilde P_\nu q_\mu}{2 m_H^2} \tilde{H}_T(q^2) \right] u(p),
\end{eqnarray}
 where $H$ =  $\Sigma$, $\Xi$ and $\Lambda$  baryons and
$j_{\mu\nu}=\bar{u}i\sigma_{\mu\nu} u-\bar{d}i\sigma_{\mu\nu} d$ is the tensor current, $\tilde P= p'+p$, 
$q = p'-p$ and $\sigma_{\mu\nu}= \frac{i}{2}[\gamma_\mu, \gamma_\nu]$ are the spin operator,
and  $u(p)$ is the spinor of the hyperon with mass $m_H$ and momentum p.

In order to calculate three tensor form factors within LCSR,
we start our analysis with the following correlation function:

\begin{equation}\label{corf}
	\Pi_{\mu\nu}(p,q)=i\int d^4 x e^{iqx} \langle 0 |T[j_{H}(0)j_{\mu \nu}(x)]|H(p)\rangle,
\end{equation}
where $J_{H}(0)$ are the hyperon interpolating fields for the $\Sigma$ and $\Xi$.
In this work, we choose the general form of the interpolating fields for $\Sigma$ and  $\Xi$  
as
 \begin{eqnarray}\label{cur}
	j_\Sigma&=&2\epsilon^{abc}\sum_{\ell=1}^{2}(u^{aT}(x) C J_1^\ell s^b(x))J_2^\ell u^c(x)\nonumber\\
	J_\Xi &=& J_\Sigma (u\leftrightarrow s) 
\end{eqnarray}
where
$J_1^1=I$, $J_1^2=J_2^1=\gamma_5$ and  $J_2^2=t$, which is an arbitrary parameter.
Choosing $t=-1$ makes the interpolating fields, which are known as Ioffe currents. 
$a$, $b$, $c$ are the color indices and $C$ is the  charge conjugation operator.
Besides, the u, d, s-quark fields are presented as $u(x)$, $d(x)$ and $s(x)$, respectively.

In order to calculate sum rules for the tensor form factors, 
we need to represent the correlation function in two different forms,
the first one is calculated in terms of quark and gluon degrees of freedom, and 
the second one is obtained using hadrons and the related correlation function 
that is written in terms of a dispersion relation.
Then, these two forms of correlation functions are equated.
In order to suppress the higher states and continuum contributions, 
we also apply the Borel transformation.

The hadronic side of the correlation function can be determined as follows,
 \begin{equation}\label{phys}
 \Pi_{\mu\nu}(p,q)= \displaystyle\sum_{p{'}} \frac{\langle0|J_H|{H(p')}\rangle\langle {H(p')}|J_{\mu \nu}|H(p)\rangle}{m^2_{H}-p'^2}+...
\end{equation}
where $m_{H}$ are the $\Sigma$, $\Xi$ and $\Lambda$ baryon mass and dots represents contributions from higher states and
continuum.
The matrix element of the interpolating current is between the vacuum and 
hyperon states, defined as
\begin{equation}\label{rezi}
 \langle0|J_H (0)|{H(p')} \rangle = \lambda_{H} u (p',s')
\end{equation}
where $\lambda_{H}$ is the hyperon overlap amplitude.
Inserting the matrix element of the tensor current in
Eq. (\ref{mat}) and the matrix element of the interpolating current in  Eq. (\ref{rezi}) into the correlation function in Eq. (\ref{phys}), we get

 \begin{eqnarray}\label{res1}
\Pi_{\mu\nu}(p,q)& =&  \frac{\lambda_H }{{m^2_{H}-p'^2}}(\pslash' +m_H )  \left[i\sigma_{\mu\nu} H_T(q^2) 
+\frac{\gamma_\mu q_\nu-\gamma_\nu q_\mu}{2 m_H} E_T(q^2)+\frac{\tilde P_\mu q_\nu-\tilde P_\nu q_\mu}{2 m_H^2} \tilde{H}_T(q^2) \right].
 \end{eqnarray}

Also, the correlation function is determined in terms of the 
quark and gluon on the QCD side.
 The interpolating fields in Eq. (\ref{cur}) are inserted into the correlation 
 function in Eq. (\ref{corf}).\\
  We obtain for $\Sigma-\Sigma$, $\Lambda-\Sigma$ and $\Xi-\Xi$ transitions,
\begin{eqnarray}\label{corrfunc}
	\Pi_{\mu\nu}&=&\frac{i}{2}\int d^4 x e^{iqx}\sum^2_{\ell=1}(C J_1^\ell)_{\alpha\beta} (J_2^\ell)_{\gamma\delta}(\sigma_{\mu \nu})_{\omega \rho}
       [\delta_\sigma^\delta \delta_\theta^\rho \delta_\phi^\beta S(-x)_{\alpha \omega}+ \delta_\sigma^\alpha \delta_\theta^\rho \delta_\phi^\beta S(-x)_{\delta \omega}]\nonumber\\
    	&& 4\epsilon^{abc}\langle 0|q_{1\sigma}^a(0) q_{2\theta}^b(x) q_{3\phi}^c(0)|H(p)\rangle
\end{eqnarray}
where $q_i$ ($i=1,~2,~3$)  denotes the quark fields, and $S(x)$ represents the quark propagator which is given as
\begin{eqnarray}\label{pro}
	S_q(x)&=&\frac{i\xslash}{2\pi^2x^4}-\frac{\langle q\bar{q}\rangle}{12}\left(1+\frac{m_0^2 x^2}{16}\right)-ig_s\int^1_0 d\upsilon\left[\frac{\xslash}{16\pi^2x^4} G_{\mu\nu}\sigma^{\mu\nu}
-\upsilon x^\mu G_{\mu\nu}\gamma^\nu\ \frac{i}{4\pi^2x^2}\right],
\end{eqnarray}
In this expression, $\langle q \bar q \rangle$ is the quark 
condensate, $m_0$ is defined in terms of the mixed quark gluon 
condensate as 
$\langle \bar q g_s G^{\mu \nu} \sigma_{\mu \nu} q\rangle \equiv  m_0^2 \langle \bar q q \rangle$ 
and $g_s$ is the strong coupling constant. $G_{\mu\nu}$ is the gluon field strength 
tensor. The terms proportional to $G_{\mu\nu}$ are expected to give 
negligibly small contributions as they are related to four and 
five-particle distribution amplitudes~\cite{Diehl:1998kh}, 
and hence  we will get neglect these terms in further analysis.
Moreover, the terms proportional to $\langle q\bar{q}\rangle$ are 
removed by Borel transformations and, finally only the first term, 
which gives the hard-quark propagator, will be considered for our discussion. 
Then, we need to know  matrix elements of the local three-quark operator,
\begin{eqnarray*}
4\epsilon^{abc}\langle 0|q_{1\sigma}^a(a_1 x) q_{2\theta}^b(a_2 x) q_{3\phi}^c(a_3 x)|H(p)\rangle
\end{eqnarray*}
where $a_1$, $a_2$ and $a_3$ are real numbers.
This matrix element can be written in terms of
distribution amplitudes (DAs) using the Lorentz covariance, 
the spin and the parity of the baryon \cite{Braun:2006hz}

\begin{eqnarray}
 &&4\epsilon^{abc}\langle 0|q_{1\sigma}^a(a_1 x) q_{2\theta}^b(a_2 x) q_{3\phi}^c(a_3 x)|H(p)\rangle 
\nonumber \\
&&= {\cal S}_1 M C_{\sigma \theta} \left(\gamma_5 H\right)_\phi +{\cal S}_2 M^2 C_{\sigma \theta} \left(\!\not\!{x} \gamma_5 H\right)_\phi 
+ {\cal P}_1 M
\left(\gamma_5 C\right)_{\sigma \theta} H_\phi + {\cal P}_2 M^2 \left(\gamma_5 C \right)_{\sigma \theta} \left(\!\not\!{x} H\right)_\phi
\nonumber \\
& &+ {\cal V}_1  \left(\!\not\!{P}C \right)_{\sigma \theta} \left(\gamma_5 H\right)_\phi + {\cal V}_2 M \left(\!\not\!{P} C \right)_{\sigma \theta}
\left(\!\not\!{x} \gamma_5 H\right)_\phi  + {\cal V}_3 M  \left(\gamma_\mu C \right)_{\sigma \theta}\left(\gamma^{\mu} \gamma_5 H\right)_\phi
\nonumber \\
&& + {\cal V}_4 M^2 \left(\!\not\!{x}C \right)_{\sigma \theta} \left(\gamma_5 H\right)_\phi+ {\cal V}_5 M^2 \left(\gamma_\mu C \right)_{\sigma \theta} \left(i
\sigma^{\mu\nu} x_\nu \gamma_5 H\right)_\phi + {\cal V}_6 M^3 \left(\!\not\!{x} C \right)_{\sigma \theta} \left(\!\not\!{x} \gamma_5 H\right)_\phi
\nonumber \\
&& + {\cal A}_1  \left(\!\not\!{P}\gamma_5 C \right)_{\sigma \theta} H_\phi + {\cal A}_2 M \left(\!\not\!{P}\gamma_5 C \right)_{\sigma \theta} \left(\!\not\!{x}
H\right)_\phi  + {\cal A}_3 M \left(\gamma_\mu \gamma_5 C \right)_{\sigma \theta}\left( \gamma^{\mu} H\right)_\phi
\nonumber \\
&& + {\cal A}_4 M^2 \left(\!\not\!{x} \gamma_5 C \right)_{\sigma \theta} H_\phi + {\cal A}_5 M^2 \left(\gamma_\mu \gamma_5 C \right)_{\sigma \theta} \left(i
\sigma^{\mu\nu} x_\nu H\right)_\phi + {\cal A}_6 M^3 \left(\!\not\!{x} \gamma_5 C \right)_{\sigma \theta} \left(\!\not\!{x} H\right)_\phi
\nonumber \\
&& + {\cal T}_1 \left(P^\nu i \sigma_{\mu\nu} C\right)_{\sigma \theta} \left(\gamma^\mu\gamma_5 H\right)_\phi + {\cal T}_2 M \left(x^\mu P^\nu i \sigma_{\mu\nu}
C\right)_{\sigma \theta} \left(\gamma_5 H\right)_\phi \nonumber \\ 
&&+ {\cal T}_3 M \left(\sigma_{\mu\nu} C\right)_{\sigma \theta}
\left(\sigma^{\mu\nu}\gamma_5 H \right)_\phi+{\cal T}_4 M \left(P^\nu \sigma_{\mu\nu} C\right)_{\sigma \theta} \left(\sigma^{\mu\varrho} x_\varrho
\gamma_5 H\right)_\phi \nonumber
\\&& + {\cal T}_5 M^2 \left(x^\nu i \sigma_{\mu\nu} C\right)_{\sigma \theta} \left(\gamma^\mu\gamma_5 H \right)_\phi 
+ {\cal T}_6 M^2 \left(x^\mu P^\nu i \sigma_{\mu\nu} C\right)_{\sigma \theta} \left(\!\not\!{x} \gamma_5
H \right)_\phi \nonumber
\\&&+ {\cal T}_7 M^2 \left(\sigma_{\mu\nu} C\right)_{\sigma \theta}
\left(\sigma^{\mu\nu} \!\not\!{x} \gamma_5 H \right)_\phi + {\cal T}_8 M^3 \left(x^\nu \sigma_{\mu\nu} C\right)_{\sigma \theta} \left(\sigma^{\mu\varrho} x_\varrho
\gamma_5 H \right)_\phi \,,\label{da-def}
\end{eqnarray}
where $H_\phi$ is the spinor of the baryon, M is the mass of the baryon, 
$C$ is the charge conjugation matrix, 
and $\sigma_{\mu\nu}=\frac{i}{2}[\gamma_\mu,\gamma_\nu]$. 
The ``calligraphic'' expressions can be expressed in terms of functions of the 
definite twist:
		    \begin{align*}
		   \mathcal{S}_1 =& S_1,\hspace{3.5cm} 2px\mathcal{S}_2=S_1-S_2,\nonumber\\
		   \mathcal{P}_1=&P_1, \hspace{3.5cm}2px\mathcal{P}_2=P_1-P_2\\
           \mathcal{V}_1=&V_1,\hspace{3.5cm} 2px\mathcal{V}_2=V_1-V_2-V_3, \nonumber\\
           2\mathcal{V}_3=&V_3,\hspace{3.5cm} 4px\mathcal{V}_4=-2V_1+V_3+V_4+2V_5,\nonumber\\
           4px\mathcal{V}_5=&V_4-V_3,\hspace{2.5cm}4(px)^2\mathcal{V}_6=-V_1+V_2+V_3+V_4 + V_5-V_6\\
           \mathcal{A}_1=&A_1, \hspace{3.5cm} 2px\mathcal{A}_2=-A_1+A_2-A_3,\nonumber\\
2\mathcal{A}_3=&A_3,\hspace{3.5cm}4px\mathcal{A}_4=-2A_1-A_3-A_4+2A_5, \nonumber\\
4px\mathcal{A}_5=&A_3-A_4, \hspace{2.5cm}4(px)^2\mathcal{A}_6=A_1-A_2+A_3+A_4-A_5+A_6\\
\mathcal{T}_1=&T_1, \hspace{3.5cm} 2px\mathcal{T}_2=T_1+T_2-2T_3, \nonumber\\
2\mathcal{T}_3=&T_7,\hspace{3.5cm} 2px\mathcal{T}_4=T_1-T_2-2T_7,\nonumber\\
2px\mathcal{T}_5=&-T_1+T_5+2T_8,\hspace{1cm}4(px)^2\mathcal{T}_6=2T_2-2T_3-2T_4+2T_5+2T_7+2T_8,
\nonumber\\ 4px \mathcal{T}_7=&T_7-T_8,\hspace{2.5cm} 4(px)^2\mathcal{T}_8=-T_1+T_2 +T_5-T_6+2T_7+2T_8
\end{align*}
where $S_i, P_i, V_i, A_i$ and $T_i$ are scalar, pesudoscalar, vector, axialvector and tensor DAs, respectively.  
The expansion of the matrix element is basically an expansion in increasing twists of the DAs. 
The twist of a DA is defined as the dimension minus the spin of the operators contributing to a given DA. 
The DAs $A_1$, $T_1$ and $V_1$ have twist three, $V_2$, $V_3$, $A_2$, $A_3$, $T_2$, $T_3$, $T_7$, $S_1$ and 
$P_1$ have twist 4, $S_2$, $P_2$, $V_4$, $V_5$, $A_4$, $A_5$, $T_4$, $T_5$ and $T_8$ are of twist 5, and $V_6$, $A_6$ and $T_6$ have twist 6. 
The DAs $F = S_i,P_i,V_i,A_i,T_i$ can be written as 
\begin{equation}
 F(a_i px) = \int dx_1 dx_2 dx_3 ~ \delta {(x_1 + x_2 + x_3 -1)}~  exp{\bigg(-ipx \sum_i {x_i a_i}\bigg)}~ F(x_i)
\end{equation}
where $x_i$ with $i = 1, 2, 3$ corresponds to longitudinal momentum fractions carried by the quarks inside the baryon. The explicit form of the 
hyperon DAs ($S_i,P_i,V_i,A_i,T_i$) is studied in detail in Refs.~\cite{Liu:2009uc, Liu:2008yg, Liu:2013bxa,Liu:2014uha}.
The DAs of the octet hyperons  up to twist-6 are investigated the basis of the QCD conformal partial wave expansion approach.
For the $\Sigma$ and $\Lambda$ baryons calculations are carried out to the next-to-leading order~\cite{ Liu:2008yg, Liu:2013bxa,Liu:2014uha} and
for the $\Xi$ baryon this calculation is carried out to the leading order of conformal spin accuracy~\cite{Liu:2009uc}.
The nonperturbative parameters (e.g. shape parameters of DAs) are obtained using the QCD sum rules~\cite{Liu:2009uc, Liu:2008yg, Liu:2013bxa,Liu:2014uha}.

Note that the hadronic representation, Eq. (\ref{res1}), and the QCD representation are obtained in different kinematical regions. The two expression can be related to each other by using the spectral representation of the correlation functions. Quite generally, the coefficients of various structures in the correlation function can be written as:
\begin{eqnarray}
\Pi(p^2,{p'}^2; Q^2) = \int_0^\infty ds_1 ds_2 \frac{\rho(s_1,s_2;Q^2)}{(s_1-p^2)(s_2-{p'}^2)} + \mbox{polynomials in $p^2$ or ${p'}^2$}
\end{eqnarray}
where $\rho$ is called the spectral density. The spectral density can be calculated both using the hadronic representation of the correlation function, $\rho^h$, or using the QCD representation, $\rho^{QCD}$. Once $\rho$ is obtained, the spectral representation allows one to evaluate the correlation function in all kinematical regions for $p^2$ and ${p'}^2$.

The LCSR are obtained by matching the expression of the correlation function in terms of QCD parameters to its expression in terms of the hadronic properties,
using their spectral representation.  In order to do this,  we choose the structures proportional to
structures $\qslash\sigma_{\mu\nu}$, $q_\mu \gamma_\nu - \gamma_\mu q_\nu$ and $q_\mu p_\nu \qslash $ for the form factors $H_T$, $E_T$ and $\tilde H_T$, respectively.
Choosing the coefficients of these structures and applying the Borel transformation with respect to the variable $p'^2 = (p+q)^2$
we obtain tensor form factors for $\Sigma - \Sigma $ transition,
\begin{eqnarray}
H_T(q^2)  \frac{\lambda_\Sigma }{{M_\Sigma
^2-p'^2}} &=&  \int_0^1 dx_2 \frac{ M_\Sigma}{(q -p x_2)^2}
 \int_0^{1-x_2}dx_1   \bigg[ P_1+T_1-T_2+T_7\bigg](x_1,x_2,1-x_1-x_2)\nonumber \\
  &&-2\int_0^1 d_\beta \frac{ M_\Sigma^3}{(q -p\beta)^4}\int_0^{\beta}d\alpha \int_\alpha^{1}dx_2 \int_0^{1-x_2}dx_1\bigg[T_1-T_2-T_5+T_6-2T_7-2T_8\bigg]\nonumber\\&&(x_1,x_2,1-x_1-x_2)\nonumber\\
\nonumber\\
E_T(q^2) \frac{\lambda_\Sigma }{{M^2_{\Sigma}-p'^2}} &=&2   \int_0^1 dx_2 \frac{ M_\Sigma}{(q -p x_2)^2}
 \int_0^{1-x_2}dx_1   \bigg[-S_1+P_1+2T_1-T_3-T_7\bigg](x_1,x_2,1-x_1-x_2)\nonumber \\
  &&-2\int_0^1 d_\beta \frac{ M_\Sigma^3}{(q -p\beta)^4}\int_0^{\beta}d\alpha \int_\alpha^{1}dx_2 \int_0^{1-x_2}dx_1\bigg[T_1-T_2-T_5+T_6-2T_7-2T_8\bigg]\nonumber\\&&(x_1,x_2,1-x_1-x_2)\nonumber\\
\nonumber\\
\tilde H_T(q^2) \frac{\lambda_\Sigma }{{M^2_{\Sigma}-p'^2}} &=&-4M_\Sigma^{3} \int_0^1 d\alpha \frac{1-\alpha}{(q -p\alpha)^4}
 \int_\alpha^{1}dx_2 \int_0^{1-x_2}dx_1\bigg[T_1-T_3-T_7\bigg](x_1,x_2,1-x_1-x_2)
\end{eqnarray}

for $\Sigma - \Lambda$ transition

\begin{eqnarray}
H_T(q^2)  \frac{\lambda_\Sigma }{{M_\Lambda
^2-p'^2}} &=&  \int_0^1 dx_2 \frac{ M_\Lambda}{(q -p x_2)^2}
 \int_0^{1-x_2}dx_1   \bigg[ P_1+T_1-T_2+T_7\bigg](x_1,x_2,1-x_1-x_2)\nonumber \\
  &&-2\int_0^1 d_\beta \frac{ M_\Lambda^3}{(q -p\beta)^4}\int_0^{\beta}d\alpha \int_\alpha^{1}dx_2 \int_0^{1-x_2}dx_1\bigg[T_1-T_2-T_5+T_6-2T_7-2T_8\bigg]\nonumber\\&&(x_1,x_2,1-x_1-x_2)\nonumber\\
\nonumber\\
E_T(q^2) \frac{\lambda_\Sigma }{{M^2_{\Lambda}-p'^2}} &=&2   \int_0^1 dx_2 \frac{ M_\Lambda}{(q -p x_2)^2}
 \int_0^{1-x_2}dx_1   \bigg[-S_1 + P_1+2T_1-T_3-T_7\bigg](x_1,x_2,1-x_1-x_2)\nonumber \\
  &&-2\int_0^1 d_\beta \frac{ M_\Lambda^3}{(q -p\beta)^4}\int_0^{\beta}d\alpha \int_\alpha^{1}dx_2 \int_0^{1-x_2}dx_1\bigg[T_1-T_2-T_5+T_6-2T_7-2T_8\bigg]\nonumber\\&&(x_1,x_2,1-x_1-x_2)\nonumber\\
\nonumber\\
\tilde H_T(q^2) \frac{\lambda_\Sigma }{{M^2_{\Lambda}-p'^2}} &=&-4M_\Lambda^{3} \int_0^1 d\alpha \frac{1-\alpha}{(q -p\alpha)^4}
 \int_\alpha^{1}dx_2 \int_0^{1-x_2}dx_1\bigg[T_1-T_3-T_7\bigg](x_1,x_2,1-x_1-x_2)
\end{eqnarray}

for $\Xi - \Xi$ transition,
\begin{eqnarray}
H_T(q^2)  \frac{\lambda_\Xi }{{m^2_{\Xi}-p'^2}} &=&  \int_0^1 dx_2 \frac{M_\Xi}{(q -p x_2)^2}
 \int_0^{1-x_2}dx_1\bigg[-V_1-V_3+A_1-A_3\bigg](x_1,x_2,1-x_1-x_2)\nonumber \\
  &&+2\int_0^1 d_\beta \frac{ M_\Xi^3}{(q -p\beta)^4}\int_0^{\beta}d\alpha \int_\alpha^{1}dx_2 \int_0^{1-x_2}dx_1
  \bigg[V_1-V_2-V_3-V_4-V_5+V_6-A_1+A_2\nonumber\\
&&-A_3-A_4+A_5-A_6\bigg](x_1,x_2,1-x_1-x_2)\nonumber\\
 \nonumber \\
E_T(q^2) \frac{\lambda_\Xi }{{m^2_{\Xi}-p'^2}} &=& 4M_\Xi \int_0^1 dx_2 \frac{1-x_2}{(q -px_2)^2}
 \int_0^{1-x_2}dx_1  \bigg[A_1-A_2-V_1+V_2 \bigg](x_1,x_2,1-x_1-x_2)\nonumber\\
 \nonumber \\
  &&+\int_0^1 d_\beta \frac{ M_\Xi^3}{(q -p\beta)^4}\int_0^{\beta}d\alpha \int_\alpha^{1}dx_2 \int_0^{1-x_2}dx_1
  \bigg[V_1-V_2-V_3-V_4-V_5+V_6-A_1\nonumber \\
  &&+A_2-A_3-A_4+A_5-A_6\bigg](x_1,x_2,1-x_1-x_2)\nonumber\\
\tilde H_T(q^2) \frac{\lambda_\Xi }{{M^2_{\Xi}-p'^2}} &=& 4M_\Xi^{3} \int_0^1 d\alpha \frac{1-\alpha}{(q -p\alpha)^4}
 \int_\alpha^{1}dx_2 \int_0^{1-x_2}dx_1 \bigg[V_1-V_2-V_3-A_1+A_2-A_3\bigg]\nonumber\\&&(x_1,x_2,1-x_1-x_2)
\end{eqnarray}
In order to eliminate the subtraction terms in the spectral representation of the correlation function,
we perform to the Borel transformation. After the transformation, contributions coming from excited and continuum states are also exponentially suppressed. Clearly, the Borel transformation and the subtraction of higher states
are achieved by using the following substitution rules (see e.g. \cite{Braun:2006hz, Braun:2001tj}):

\begin{eqnarray}
		&&\int dx \frac{\rho(x)}{(q-xp)^2}\rightarrow -\int_{x_0}^1\frac{dx}{x}\rho(x) e^{-s(x)/M^2}, \nonumber		\\
		&&\int dx \frac{\rho(x)}{(q-xp)^4}\rightarrow \frac{1}{M^2} \int_{x_0}^1\frac{dx}{x^2}\rho(x) e^{-s(x)/M^2}+\frac{\rho(x_0)}{Q^2+x_0^2 m_H^2} e^{-s_0/M^2},\nonumber\\
\end{eqnarray}
where,
\begin{eqnarray}
s(x)=(1-x)m_H^2+\frac{1-x}{x}Q^2,
\end{eqnarray}
$M$ is the Borel mass  and $x_0$ is the solution of the quadratic equation for $s=s_0$:
\begin{eqnarray}
x_0=\frac{\sqrt{(Q^2+s_0-m_H^2)^2+4m_H^2 Q^2}-(Q^2+s_0-m_H^2)}{2m_H^2},
\end{eqnarray}
where $s_0$ is the continuum threshold.


\section{Results and Discussion}

\hspace*{0.5 cm}In this section, we present the numerical results of the octet-octet hyperon isovector tensor transition form factors.
In this work, we use the hyperon DAs which depend on various 
nonperturbative parameters such as,  $f_{\Sigma}$, $f_\Xi$ and 
$f_ {\Lambda}$
~\cite{Liu:2009uc, Liu:2008yg, Liu:2013bxa,Liu:2014uha}.
 The DAs of the $\Sigma$, $\Xi$ and $\Lambda$ baryons have been calculated  by employing the
 QCD sum rules without higher-order terms in Ref.~\cite{Liu:2009uc, Liu:2008yg}.
 Then the study including higher-order corrections have been done for
  $\Sigma$ and $\Lambda$ baryon in Ref.~\cite{Liu:2013bxa, Liu:2014uha}. 
In Table \ref{parameter_table}, we present the values of the input parameters using the
DAs of $\Sigma$, $\Xi$ and $\Lambda$ baryons. 
For the numerical analysis, we use the values of the hyperon masses as follow; 
$M_\Lambda = 1.11~GeV$, $M_\Sigma = 1.2~GeV$,
 $M_\Xi = 1.3~GeV$ \cite{Beringer:1900zz}. 
Besides, we need also specify the values of the residues of , $\Sigma$ and $\Xi$  baryon. The residues can be determined from the mass sum rules as   $\lambda_\Sigma = 0.039~GeV^3$ and $\lambda_\Xi = 0.040~GeV^3$ for $\Sigma$ and $\Xi$, respectively~\cite{Aliev:2002ra}.

\begin{table}[t]
	\addtolength{\tabcolsep}{2pt}
	\begin{center}
\begin{tabular}{ |l|l|l| }
\hline
\multicolumn{3}{ |c| }{DAs Parameters} \\
\hline\hline
$~~~~~~~~~~~~~~~~~~~~\Sigma$ & $~~~~~~~~~~~~~~~~~~~~\Lambda $& $~~~~~~~~~~~~~~~~~~~~\Xi$ \\ \hline\hline
\multirow{4}{*}{} &  &  \\
 $f = (9.4 \pm 0.4)\times 10^{-3}$~GeV$^2$        & $f  = (6.0 \pm 0.3) \times 10^{-3}~GeV^2$        & $f  = (9.9 \pm 0.4) \times10^{-3}~GeV^2$ \\
 $\lambda_1= (-2.5 \pm 0.1)\times  10^{-2}$~GeV$^2$  &$\lambda_1 = (1.0 \pm 0.3) \times 10^{-2}~GeV^2$   & $\lambda_1 = (-2.8 \pm 0.1) \times10^{-2}~GeV^2$\\
 $\lambda_2 = (4.4 \pm 0.1) \times  10^{-2}$~GeV$^2$  & $\lambda_2 = (0.83\pm 0.05) \times10^{-2}~GeV^2$  & $\lambda_2 = (5.2 \pm 0.2)10^{-2} \times ~GeV^2$\\ $\lambda_3 = (2.0  \pm 0.1) \times  10^{-2}$~GeV$^2$  
 & $\lambda_3 = (0.83\pm 0.05) \times 10^{-2}~GeV^2$  & $\lambda_3 = (01.7 \pm 0.1) \times 10^{-2}~GeV^2$
\\
 $V_1^s = 0.39 \pm 0.01$   & $A_1^s = 0.31 \pm 0.01$ &  \\
 $A_1^u = 0.29 \pm 0.12$   & $A_1^q = 0.032 \pm 0.006$ &  \\
 $f_1^s = -0.15 \pm 0.12$  & $f_1^s = 0.23 \pm 0.01$ &  \\
 $f_2^s = 9.9 \pm 2.5$     & $f_1^q = -0.23 \pm 0.03$ &  \\
 $f_3^s = 1.6 \pm 0.2$     & $f_3^q = 0.43 \pm 0.07$ &  \\
 $f_1^u = -0.11 \pm 0.01$ & $f_4^q = 1.07 \pm 0.12$ &  \\ 
 $P_2^0 = 0.004 \pm 0.0004$ &  &  \\
 $S_1^u = -0.0014 \pm 0.0002$ &  &  \\ \hline\hline
\end{tabular}
\caption{The values of the parameters are used in the DAs of  $\Sigma$, $\Lambda$  and $\Xi$.
          }
          \label{parameter_table}
         \end{center}
\end{table}

In the traditional analysis of sum rules, the spectral density of
the higher states and the continuum are parameterized using quark hadron duality. In this approach,  the spectral density corresponding to the contributions of the higher states and continuum is parameterized as
$$\rho^h(s) = \rho^{QCD}(s) \theta(s - s_0).$$
The predictions for  the form factors depend on two auxiliary parameters:
the squared of Borel mass $M^2$, and the continuum threshold $s_0$.
The continuum threshold signals the scale at which, the excited states and
continuum start to contribute to the correlation function. Hence it is expected that $s_0\simeq(m_\Sigma+0.3)^2 ~GeV^2=2.25~GeV^2$,  $s_0\simeq(m_\Lambda+0.3)^2~GeV^2 \simeq 1.99~GeV^2$ and $s_0\simeq(m_\Xi+0.3)^2~GeV^2=2.56~GeV^2$.
One approach to determine the continuum threshold and the working region of the Borel
parameter $M^2$ is to plot the dependence of the predictions on $M^2$ for a range of values
of the continuum threshold and determine the values of $s_0$ for which there is a stable
region with respect to variations of the Borel parameter $M^2$. For this reason,
in Figs.(1)-(3), we plot the dependence of the form factors on $M^{2}$ for  fixed values of $Q^2$ and various values of $s_0$ in the region $2 ~ GeV^2 \leq s_0 \leq 4~GeV^2$.
 As can be seen from  figures (in the case of  old DAs and new DAs), for $s_0=2.5\pm0.5~GeV^2$, the results are practically
independent of the value of $M^2$ for the shown range.
The uncertainty due to variations of $s_0$ in this range is much larger than the uncertainty due to variations with respect to $M^2$. Note that the determined range of $s_0$ is in the range that one would expect from the physical interpretation of $s_0$.

In Figs. (4)-(6), we present the $Q^2$ dependence of the form factors obtained using two DAs. Our observations can be summarized as follows;
\begin{enumerate}
\item $\Sigma-\Sigma$ case:

We show the behaviour of the form factors that agree well with 
our expectations. 
The values of the tensor form factors  decrease quickly as 
we increase the momentum transfer. 
In Fig. (4-a, c and e) are represent the results of the 
new DAs and Fig. (4-b, d and f) are represent the results of old DAs. 
In both case, the $Q^2$ dependence of the form factors are 
similar but the values of the new DAs results are larger than 
old DAs results. 
We see that higher-order terms gives dominant contribution.

\item $\Sigma-\Lambda$ case: 

In Fig. (5-a, c and e) are represent the results of the new DAs 
and Fig. (5-b, d and f) are represent the results of old DAs.
In the case of old DAs, the results of the form factors 
 $E_T^{\Sigma \Lambda}$ in 
 negatif region but new DAs result change the behaviour this 
 form factor. 
In the case of $H_T^{\Sigma \Lambda}$,  the $Q^2$ dependence of 
the form factor is similar behaviour and stable but the values of 
the new DAs results are larger than old one.
 In the case of old DAs, the results of the form factor 
 $\tilde H_T^{\Sigma \Lambda}$ in 
 positive region but new DAs result change the behaviour this 
 form factor.

\item $\Xi-\Xi$ case

The higher-order terms of  DAs of $\Xi$ baryon have not yet been calculated. So
there is only one DAs result showing the behavior of the form factors 
that agree well with our expectations. The values of the tensor form factors  
decrease quickly as we increase the momentum transfers.
\end{enumerate}

Unlike other form factors the tensor form factors are renormalization-scale dependent \cite{He:1994gz}.
The numerical values of DAs are used at the scale $\mu^2 = 1$ $ GeV^2$ in Ref. \cite{Chernyak:1984bm}, therefore, in present work our predictions correspond to this scale. In order to compare our results, we use the following expressions ~\cite{Barone:2001sp}:
\begin{equation} \label{RGE}
	F(\mu^2)=\left(\alpha_s(\mu^2)\over \alpha_s(\mu_i^2)\right)^\frac{4}{33-2n_f}
	\left[1-\frac{337}{468\pi}[\alpha_s(\mu_i^2)-\alpha_s(\mu^2)]\right]F(\mu_i^2),
\end{equation}
where $n_f$ is the number of flavors, $\mu_i$ is the initial renormalization scale and
\begin{align}
	\alpha_s(\mu^2)=\frac{4\pi}{9\ln(\mu^2/\Lambda^2)}\left[1-
	\frac{64}{81}\frac{\ln(\ln(\mu^2/\Lambda^2))}{\ln(\mu^2/\Lambda^2)}\right].
\end{align}

 The values of the form factors at zero momentum transfer, $Q^2 = 0$, defines the corresponding charges.
 However, in our case, the working region
of the LCSR cannot extrapolate to the $Q^2 = 0$ directly. LCSR results more reliable at $Q^2 > 1~GeV^2$.
  The tensor form factor is parameterized in terms of an exponential form
\begin{equation}
	F_{T}(Q^2)= F_T(0) \exp[-Q^2/m_{T}^2]
\end{equation}
which makes a reasonable description of data with a two-parameter fit.
Our predictions are presented in Table 2. 
As seen in Table 2,  predictions of old DAs are more reliable and reasonable 
than the new DA result. The results of new DAs are very large and quite suspicious. 
When one takes the nonrelativistic limit, the isovector tensor charge becomes 
identical with the isovector axial vector charge \cite{Jaffe:1991kp}, 
which is of order similar to the hyperon isovector axial 
vector charges [ $g_A^\Sigma \simeq 1$, 
$g_A^\Xi \simeq 0.3$  \cite{Erkol:2011qh, Erkol:2009ev}]. 
Therefore obtained by using new DAs results unreliable.
Some of the results obtained using the old DAs are consistent  
but some of them are not consistent with this prediction.

We give the results for the  $H_T^\Sigma = 1.10$ and $H_T^\Xi = -0.30$ obtained from chiral quark soliton model \cite{Ledwig:2010tu}, which have been calculated at a renormalization scale  $\mu = 0.36~GeV^2$.  In order to compare our results, we use the   Eq. (16) to the relate this results of form factors to those at  $\mu = 1~GeV^2$, and we obtained these result;   $H_T^\Sigma(0) = 1.00$ and $H_T^\Xi(0) = -0.27$. 
As seen from Table 2, our results are different from the results 
obtained in the chiral quark soliton model.

\begin{table}
\hspace*{-0.5cm}
\begin{tabular}{ |l|l|l|l|l|l|l|l}
\hline
&\multicolumn{2}{|c|}{Results of old DAs} &\multicolumn{2}{|c|}{Results of new DAs}\\
\hline
\begin{tabular}{c}Transition \end{tabular}& \begin{tabular}{c} $F_T(0)(GeV^{-2})$ \end{tabular} & $m_T(GeV)$& $F_T(0)(GeV^{-2})$ & $m_T(GeV)$ \\ \hline\hline
\multirow{2}{*} &$E_T(0)= -0.38 \pm 0.04 $&$1.31 $&$E_T(0)=-115.39 \pm 15.44$ &$1.28$   \\
{$\Sigma-\Sigma$}&$H_T(0)=-0.59 \pm 0.11$ &$1.30$ &$H_T(0)=-137.72 \pm 30.74$ &0.96  \\
&$\tilde H_T(0)= -0.10\pm 0.01$ &$1.10$ &$\tilde H_T(0)=-19.28\pm 2.50$ &1.48   \\
\hline
\multirow{2}{*} &$E_T(0)= -0.6 \pm 0.08$ & $1.93$&$E_T(0)=2160.45 \pm 120.44$&$1.26$  \\
{$\Sigma-\Lambda$}&$H_T(0)= 0.65 \pm 0.1$ &$1.74$ &$H_T(0)=135.14 \pm 30.14$4 &$1.26$   \\
&$\tilde H_T(0)= -0.001 \pm 0.0001$ &$1.65$ &$\tilde H_T(0)=-14.76 \pm 0.50$ &$1.28$  \\
\hline
\multirow{2}{*} &$E_T(0)=1.09 \pm 0.2$& $1.32$&$-$&$-$ \\
{$\Xi-\Xi$}&$H_T(0)=-3.00 \pm 0.6$ &$1.23$ &$- $&$ -$  \\
&$\tilde H_T(0)=-0.10\pm 0.01$ &$1.22$ &$- $&$-$\\
\hline \hline
\end{tabular}
\caption{The values of exponential fit parameters, $F_T(0)$ and $m_{T}$ for tensor  form factor obtained from the old and new DAs analysis of sum rules.}
	\label{fit_table}
\end{table}

In conclusion, we have evaluated the isovector tensor form factors of 
octet-octet hyperons by applying the LCSR. 
These form factors are related to the transverse polarization which  
gives an important piece of information on the internal structure of 
baryons (e.g. the transverse spin structure of the baryons).  
The $Q^2$ dependency of form factors are obtained using the old and new DA results. 
The new DA results shows that our predictions of the form factors are very large. 
Our predictions on the isovector  tensor charges can be summarized in Table 2. 
The old DA results seems to be more reliable. 
Our results on these form factors are compared with the chiral quark 
soliton model predictions. The chiral quark soliton  model results only 
exist for the $H_T$ form factor so  we cannot compare results of 
other form factors.

\section*{Acknowledgments}
This work has been supported,
 by the Scientific and Technological Research Council of Turkey 
 (TUBITAK) under project No. 114F278.


\newpage

\begin{figure}[htp]
\centering
 \subfloat[]{\label{fig:SigEtMsq.eps}\includegraphics[width=0.35\textwidth]{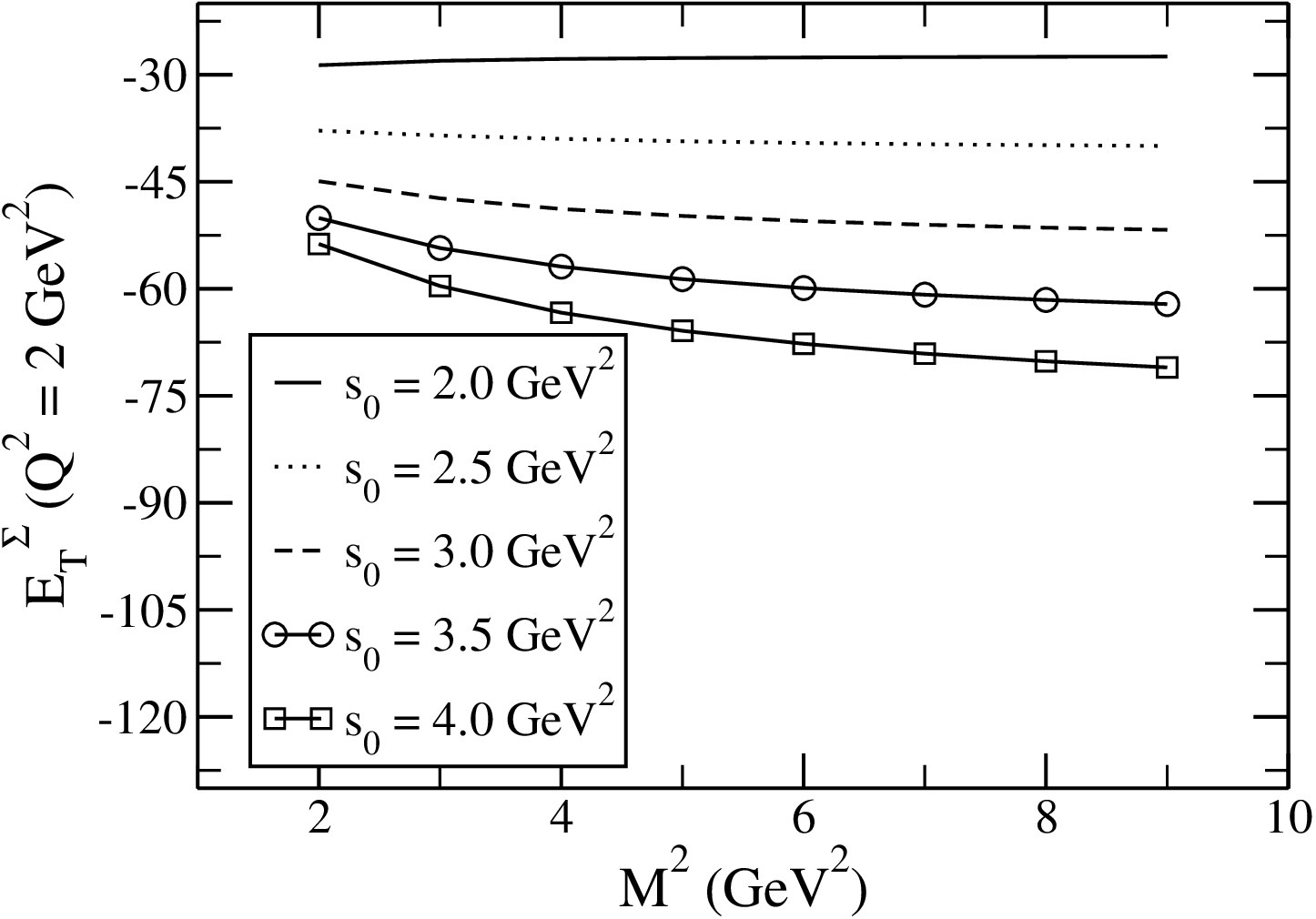}}
 \subfloat[]{\label{fig:OSigEtMsq.eps}\includegraphics[width=0.35\textwidth]{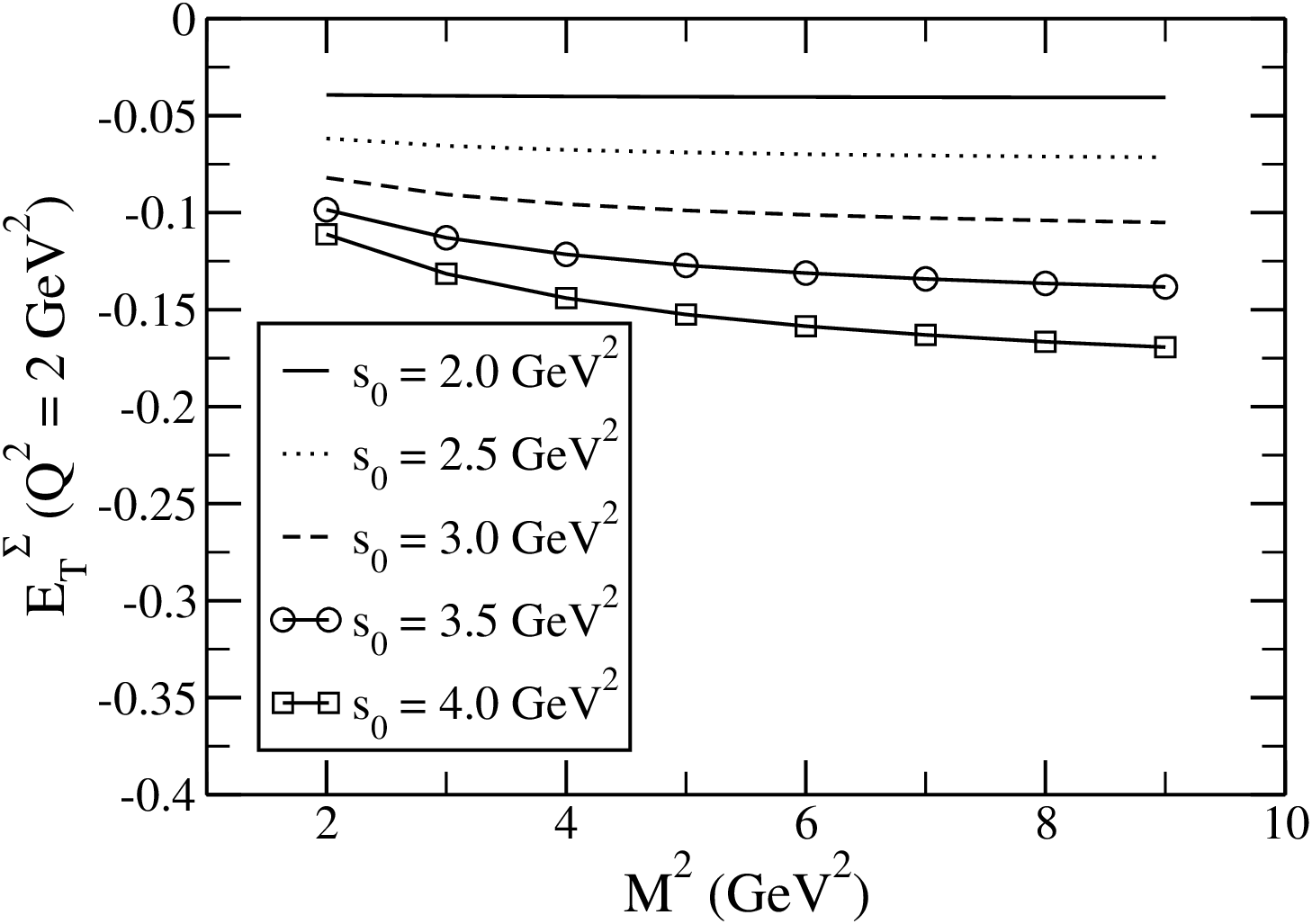}}\\
  \subfloat[]{\label{fig:SigHtMsq.eps}\includegraphics[width=0.35\textwidth]{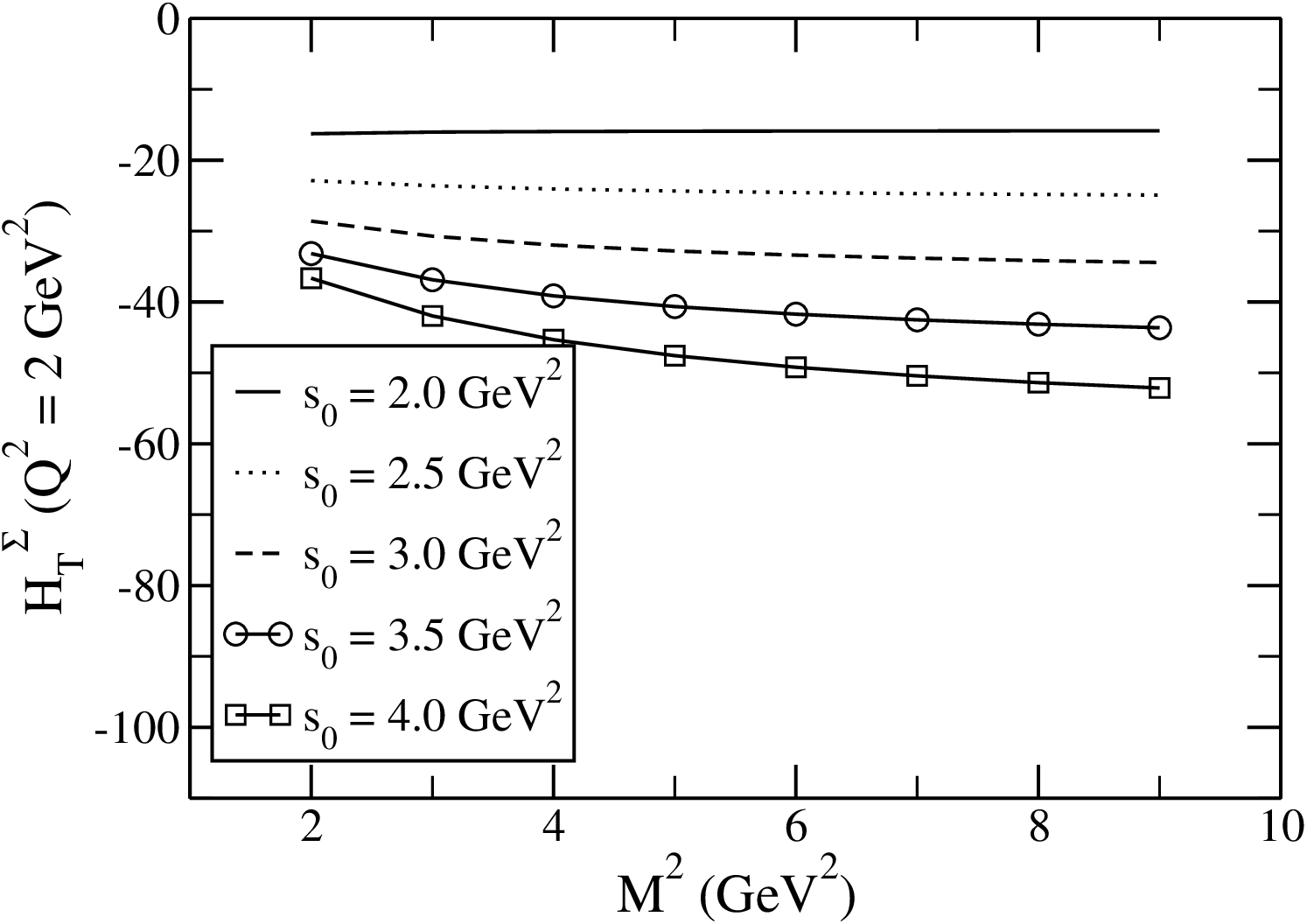}}
  \subfloat[]{\label{fig:OSigHtMsq.eps}\includegraphics[width=0.35\textwidth]{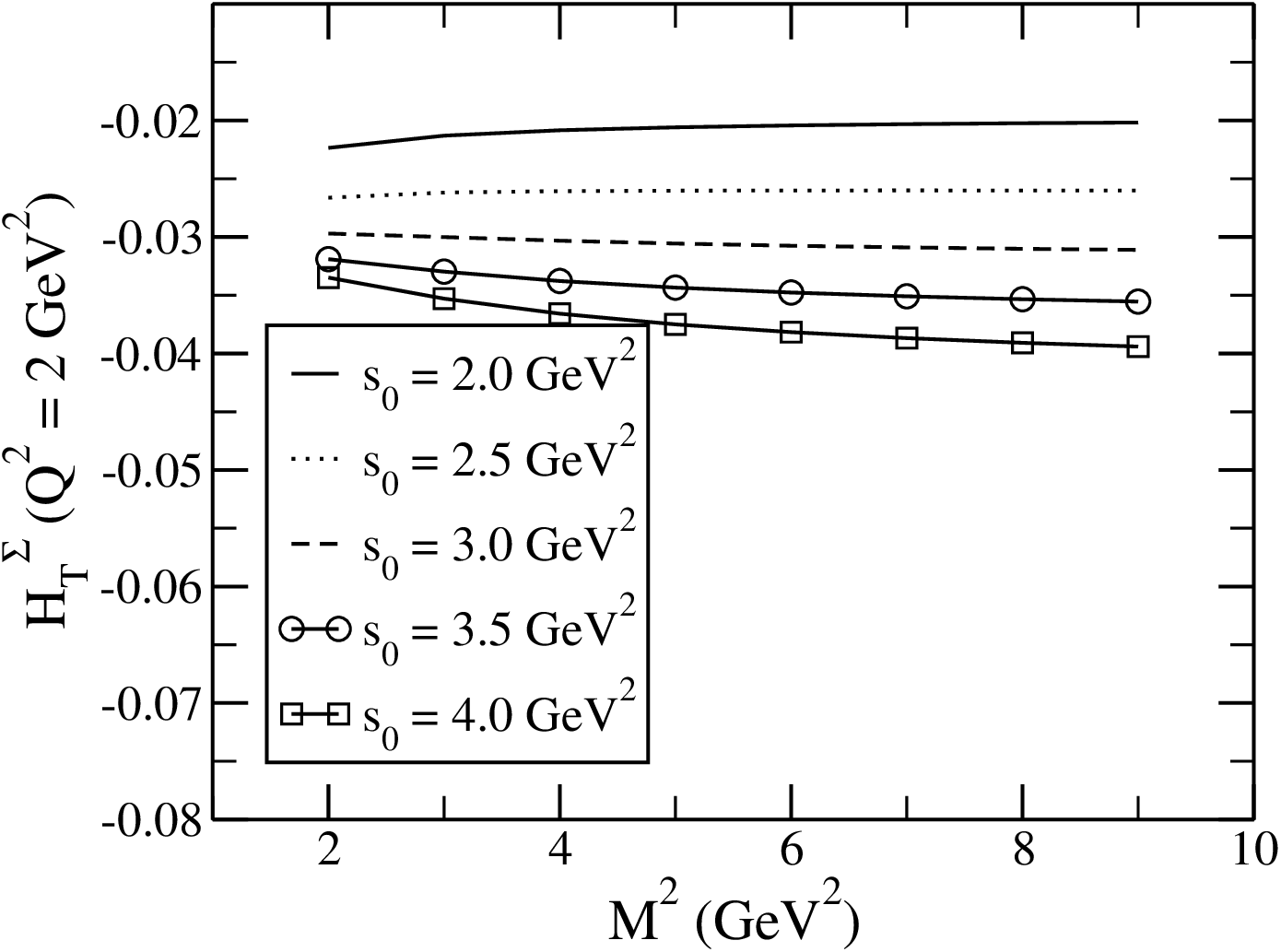}}\\
   \subfloat[]{\label{fig:SigHtilMsq.eps}\includegraphics[width=0.35\textwidth]{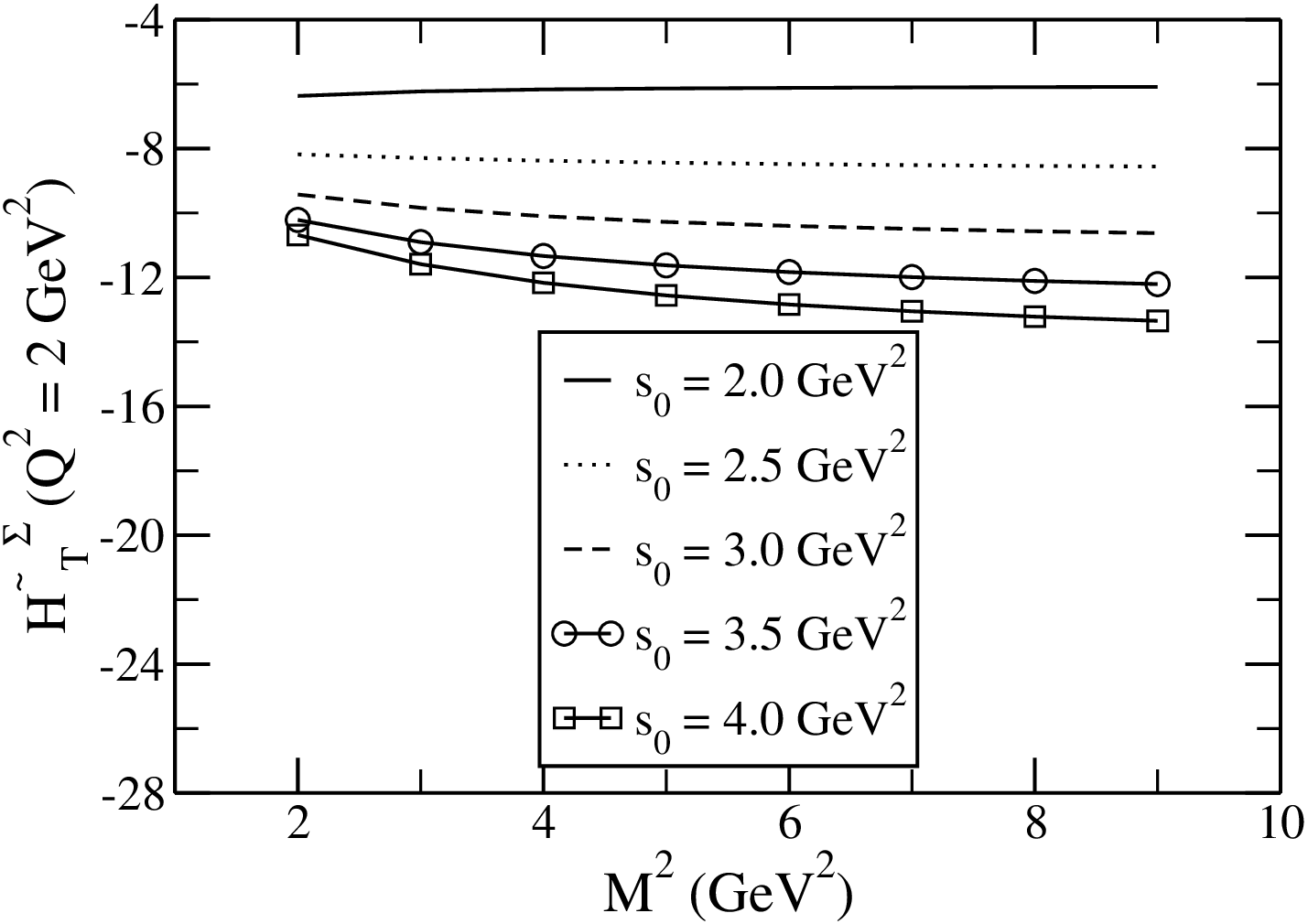}}
   \subfloat[]{\label{fig:OSigHtilMsq.eps}\includegraphics[width=0.35\textwidth]{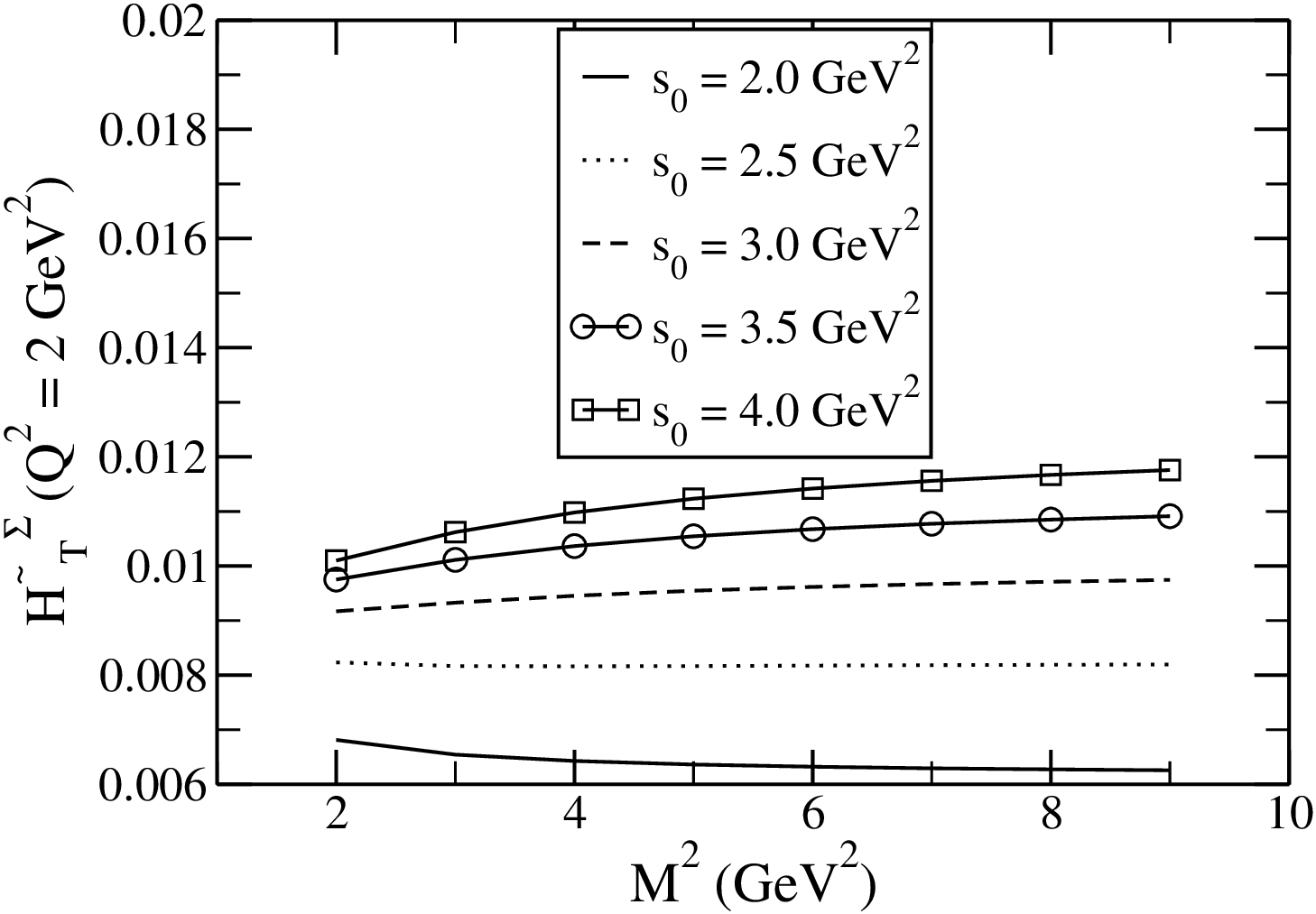}}
  \caption{  The dependence of the form factors; on the Borel parameter squared $M^{2}$
  for the values of the continuum threshold $s_0 = 2 ~GeV^2$, $s_0 = 2.5~GeV^2$,  $s_0 = 3 ~GeV^2$,
  $s_0 = 3.5~GeV^2$ and $s_0 = 4~GeV^2$ and $Q^2 = 2~GeV^2$,
    (a) and (b) for $E_T^\Sigma$ tensor form factor,
    (c) and (d) for $H_T^\Sigma$ tensor form factor,
    (e) and (f) for $\tilde H_T^\Sigma$ tensor form factor. In here (a), (c) and (e) are represent results of new DAs and (b), (d) and (f) are represent result of old DAs.}
\end{figure}

\begin{figure}[htp]
\centering
 \subfloat[]{\label{fig:lamSEtMsq.eps}\includegraphics[width=0.35\textwidth]{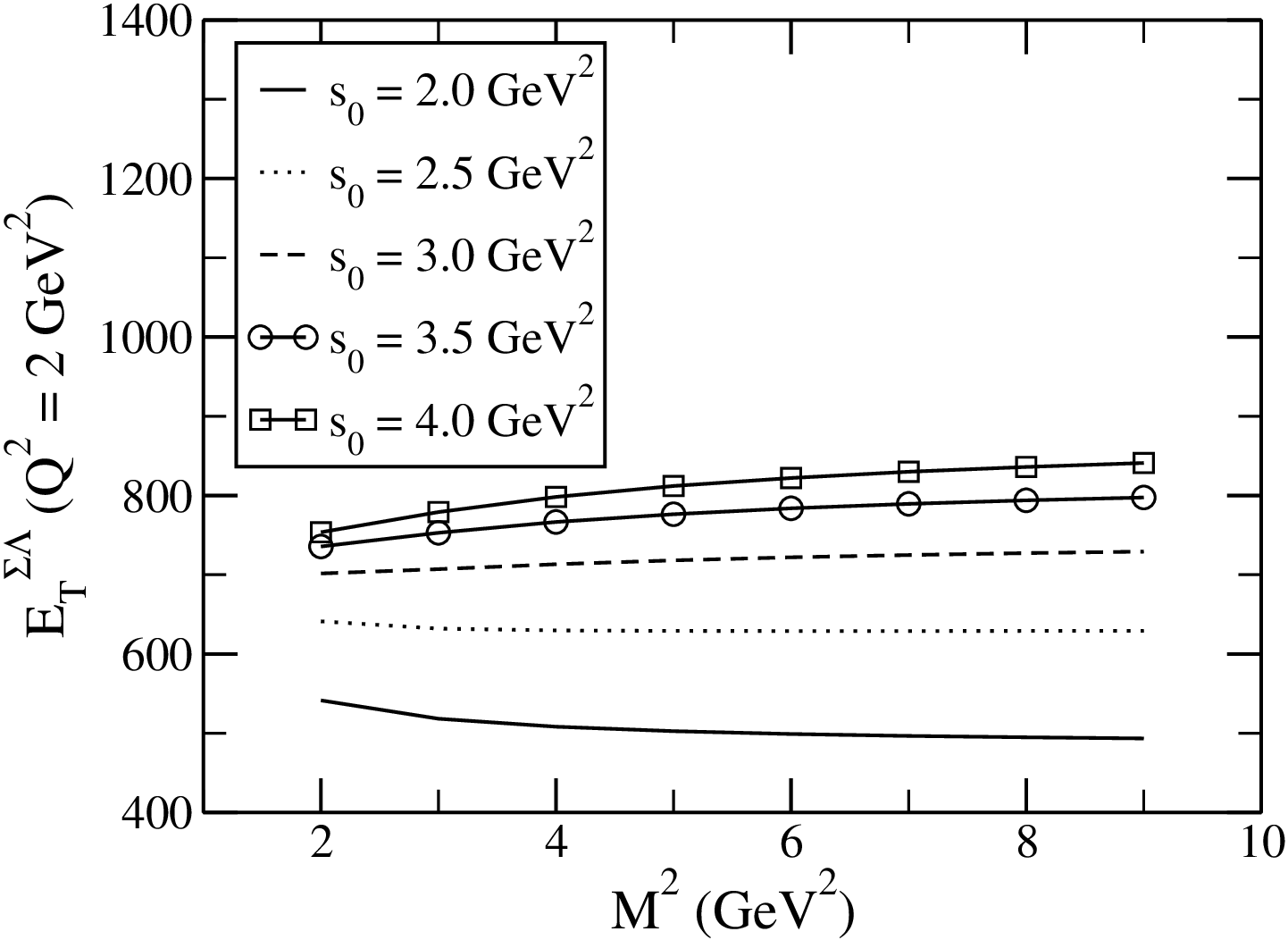}}
 \subfloat[]{\label{fig:OlamStMsq.eps}\includegraphics[width=0.35\textwidth]{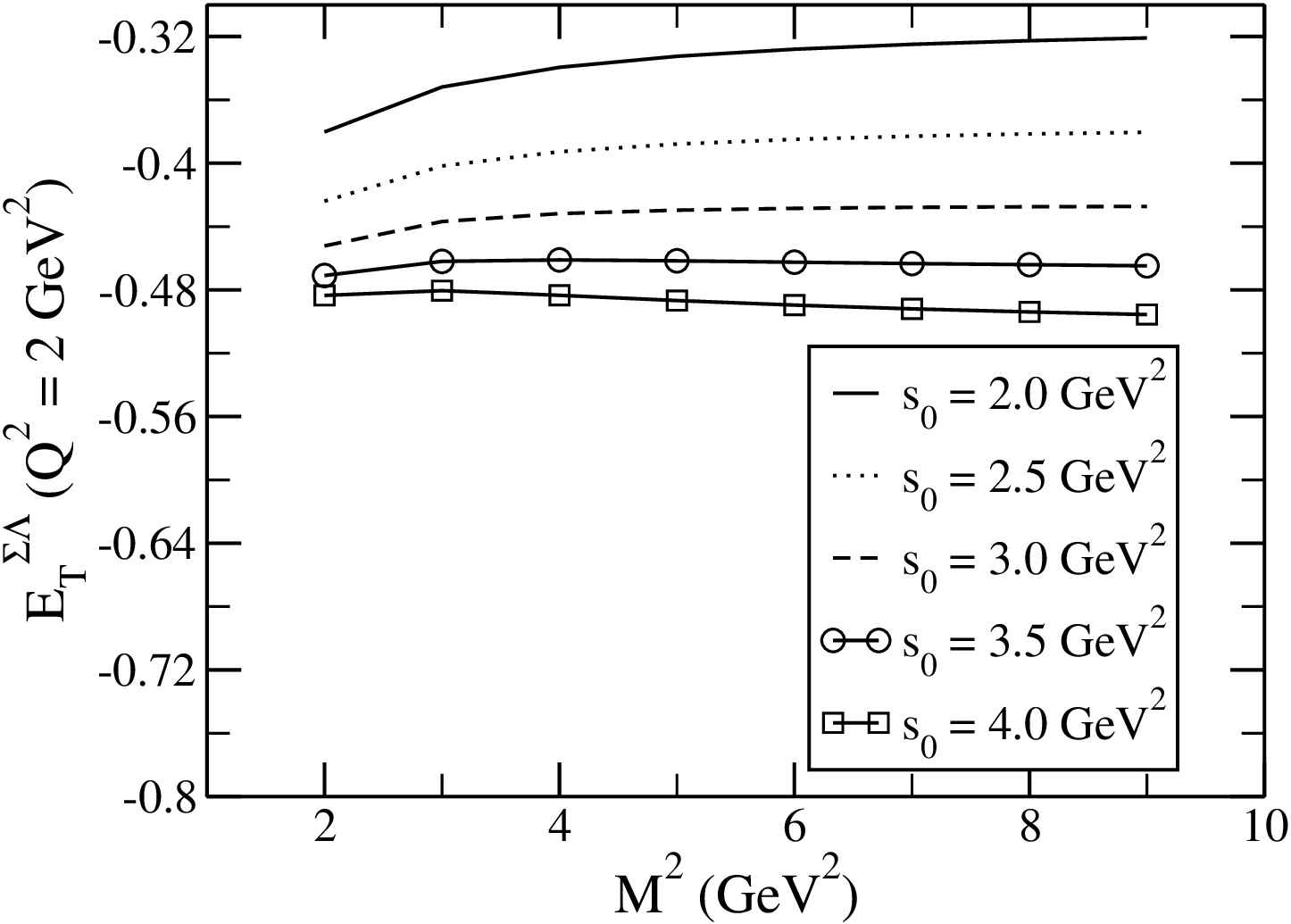}}\\
  \subfloat[]{\label{fig:lamSHtMsq.eps}\includegraphics[width=0.35\textwidth]{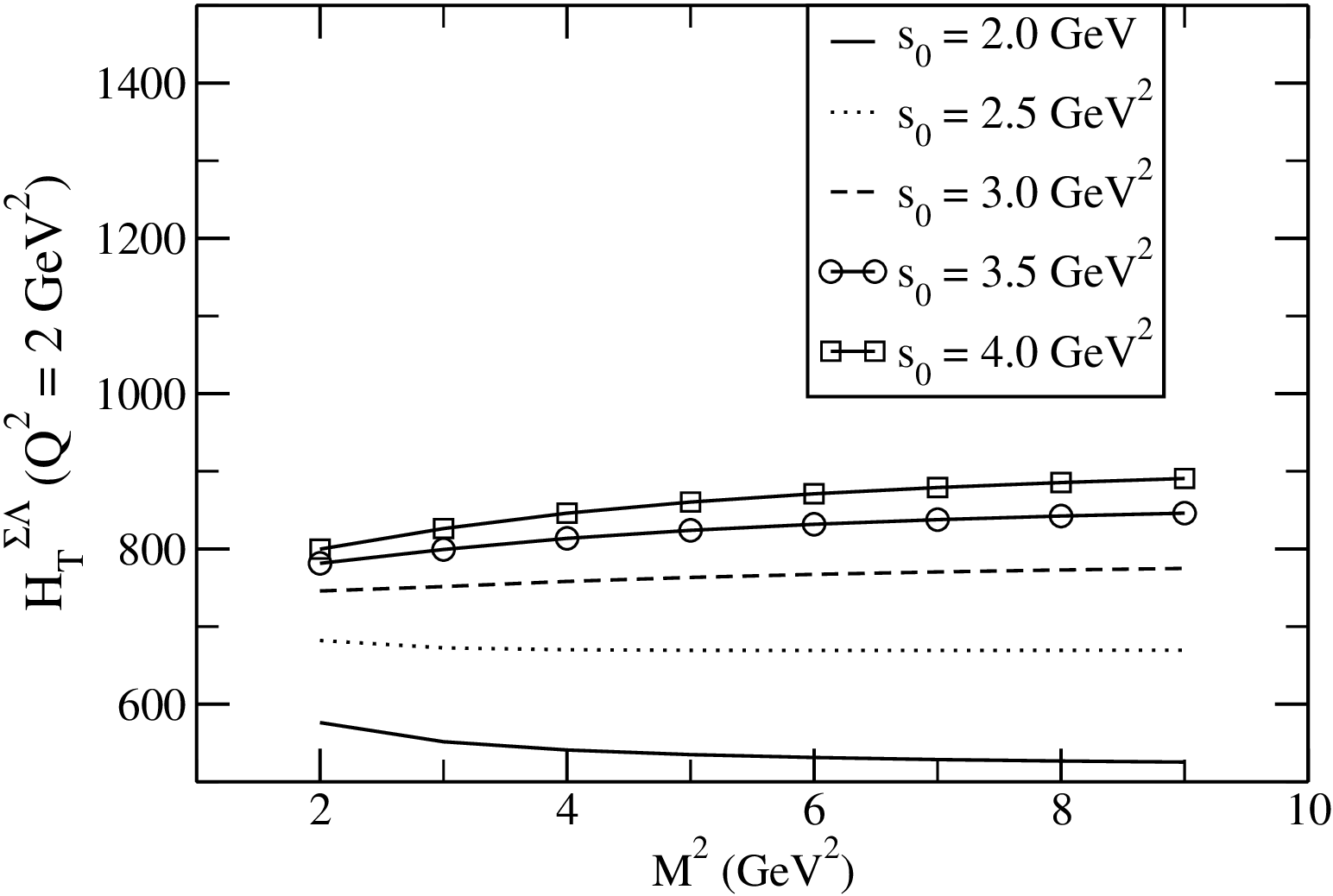}}
  \subfloat[]{\label{fig:OlamSHtMsq.eps}\includegraphics[width=0.35\textwidth]{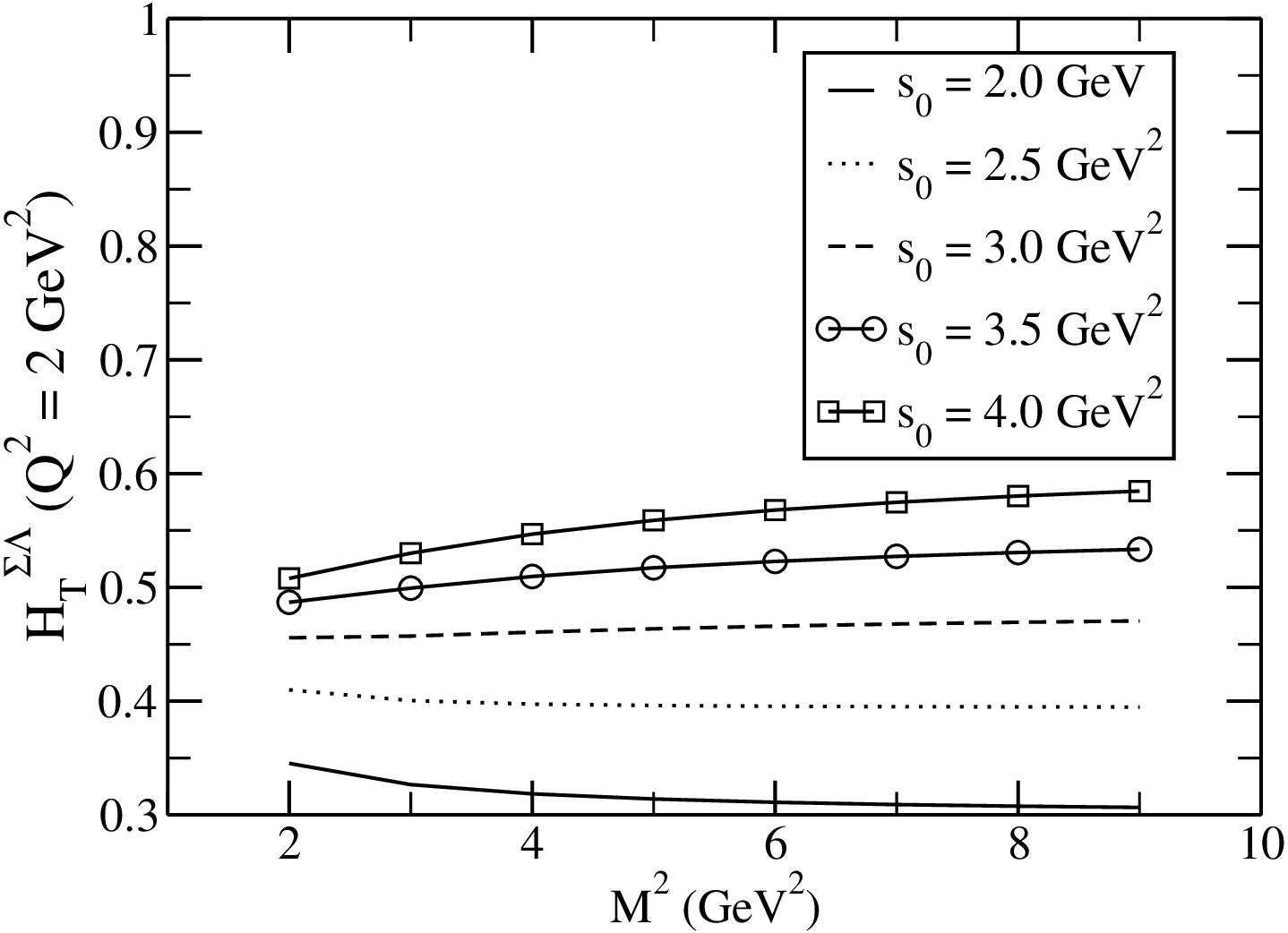}}\\
  \subfloat[]{\label{fig:lamSHtilMsq.eps}\includegraphics[width=0.35\textwidth]{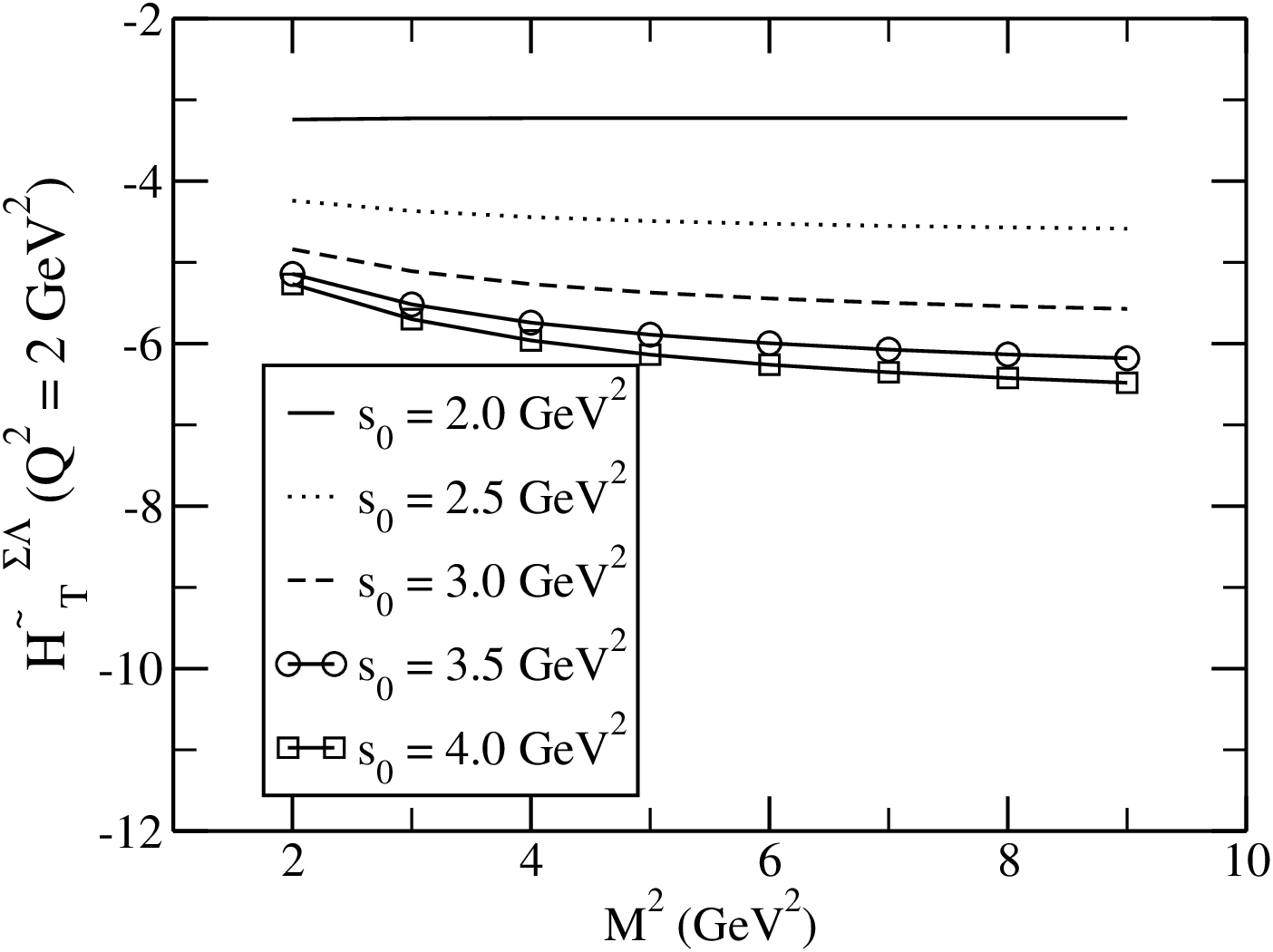}}
  \subfloat[]{\label{fig:OlamSHtilMsq.eps}\includegraphics[width=0.35\textwidth]{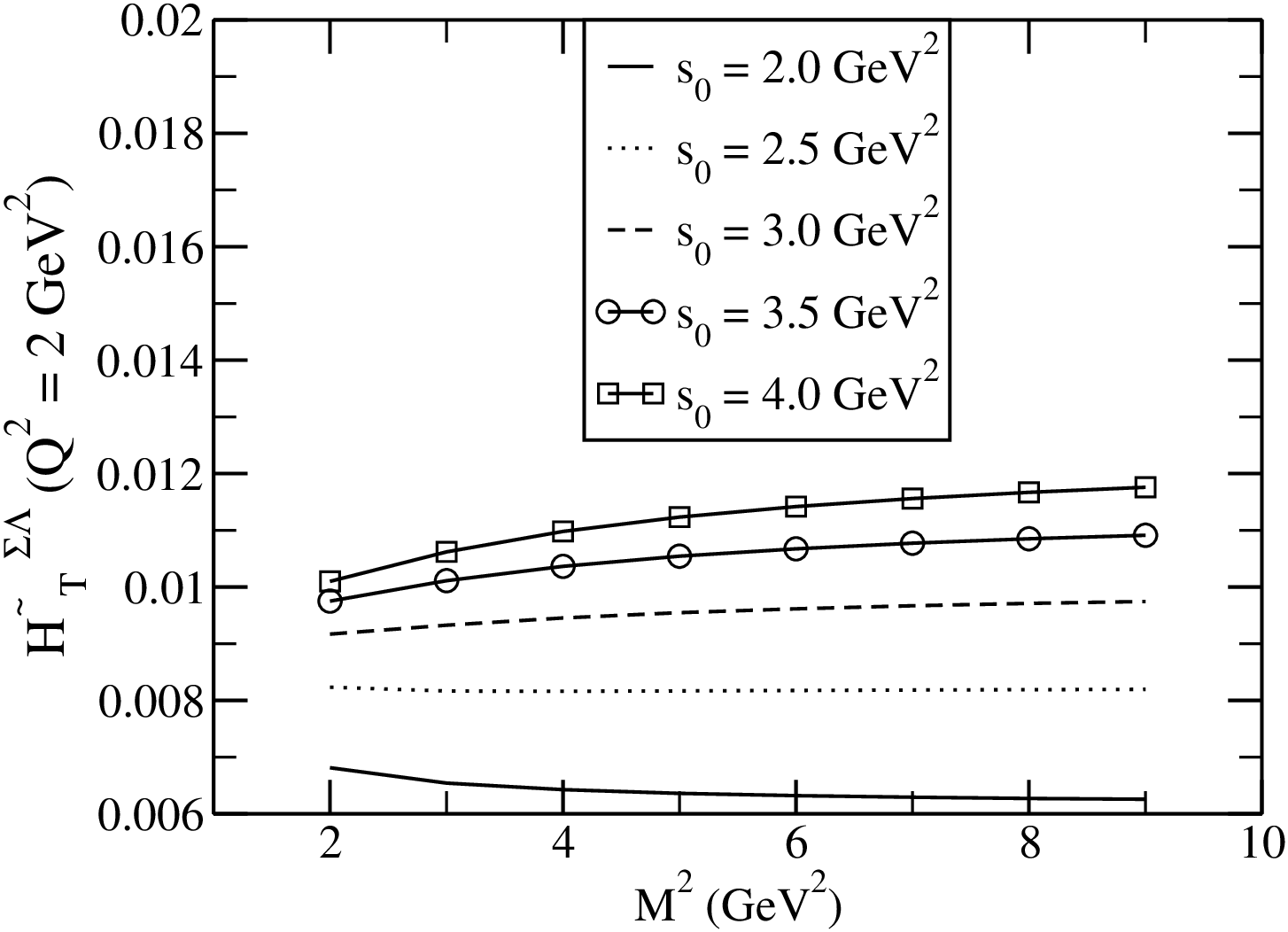}}
  \caption{ The dependence of the form factors; on the Borel parameter squared $M^{2}$
  for the values of the continuum threshold $s_0 = 2 ~GeV^2$, $s_0 = 2.5~GeV^2$,  $s_0 = 3 ~GeV^2$,
  $s_0 = 3.5~GeV^2$ and $s_0 = 4~GeV^2$ and $Q^2 = 2~GeV^2$,
    (a) and (b) for $E_T^{\Sigma \Lambda}$ tensor form factor,
    (c) and (d) for $H_T^{\Sigma \Lambda}$ tensor form factor,
    (e) and (f) for $\tilde H_T^{\Sigma \Lambda}$ tensor form factor.  In here (a), (c) and (e) are represent results of new DAs and (b), (d) and (f) are represent result of old DAs.}
\end{figure}

\begin{figure}[htp]
\centering
 \subfloat[]{\label{fig:ChiEtMsq.eps}\includegraphics[width=0.35\textwidth]{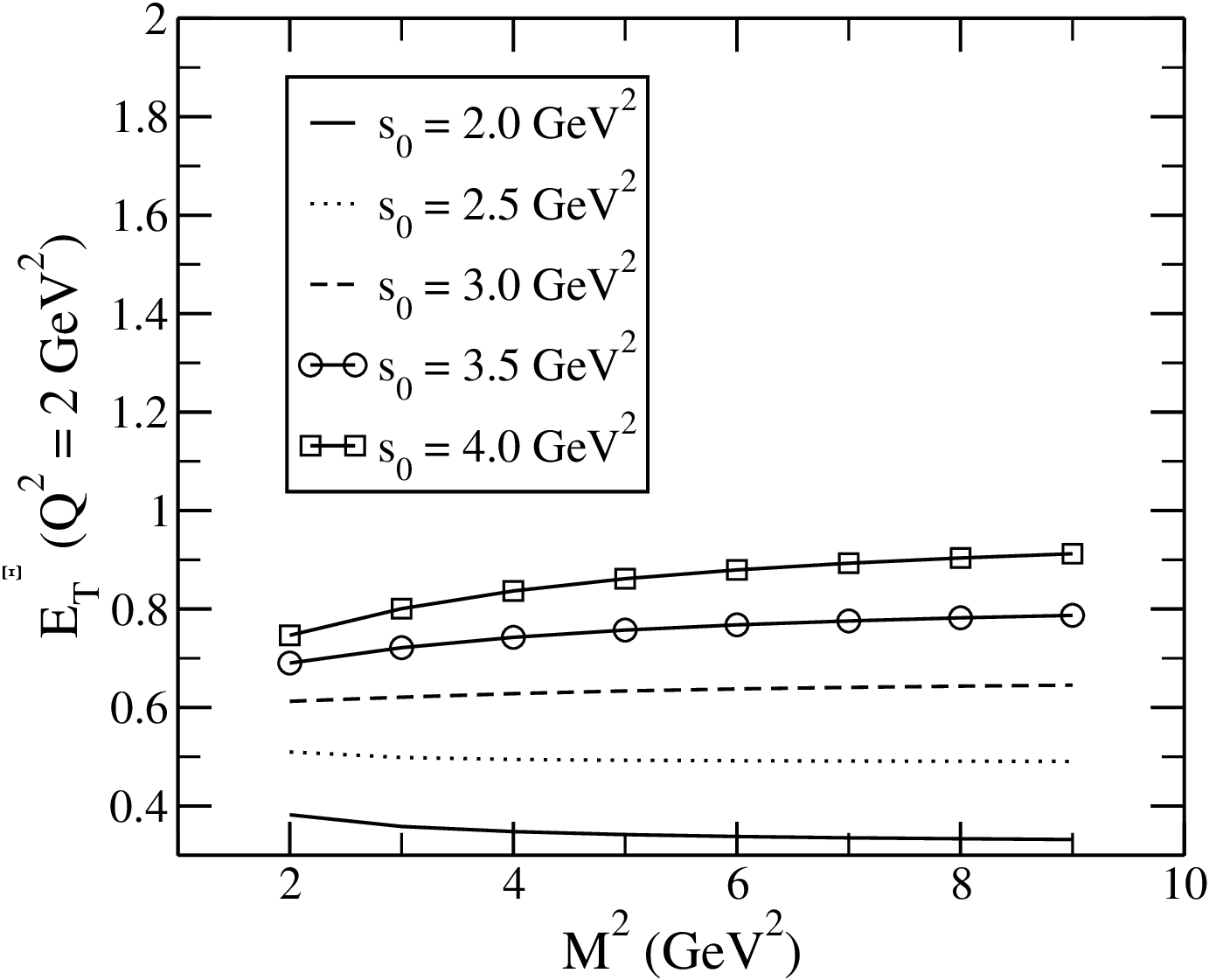}}\\
  \subfloat[]{\label{fig:ChiHtMsq.eps}\includegraphics[width=0.35\textwidth]{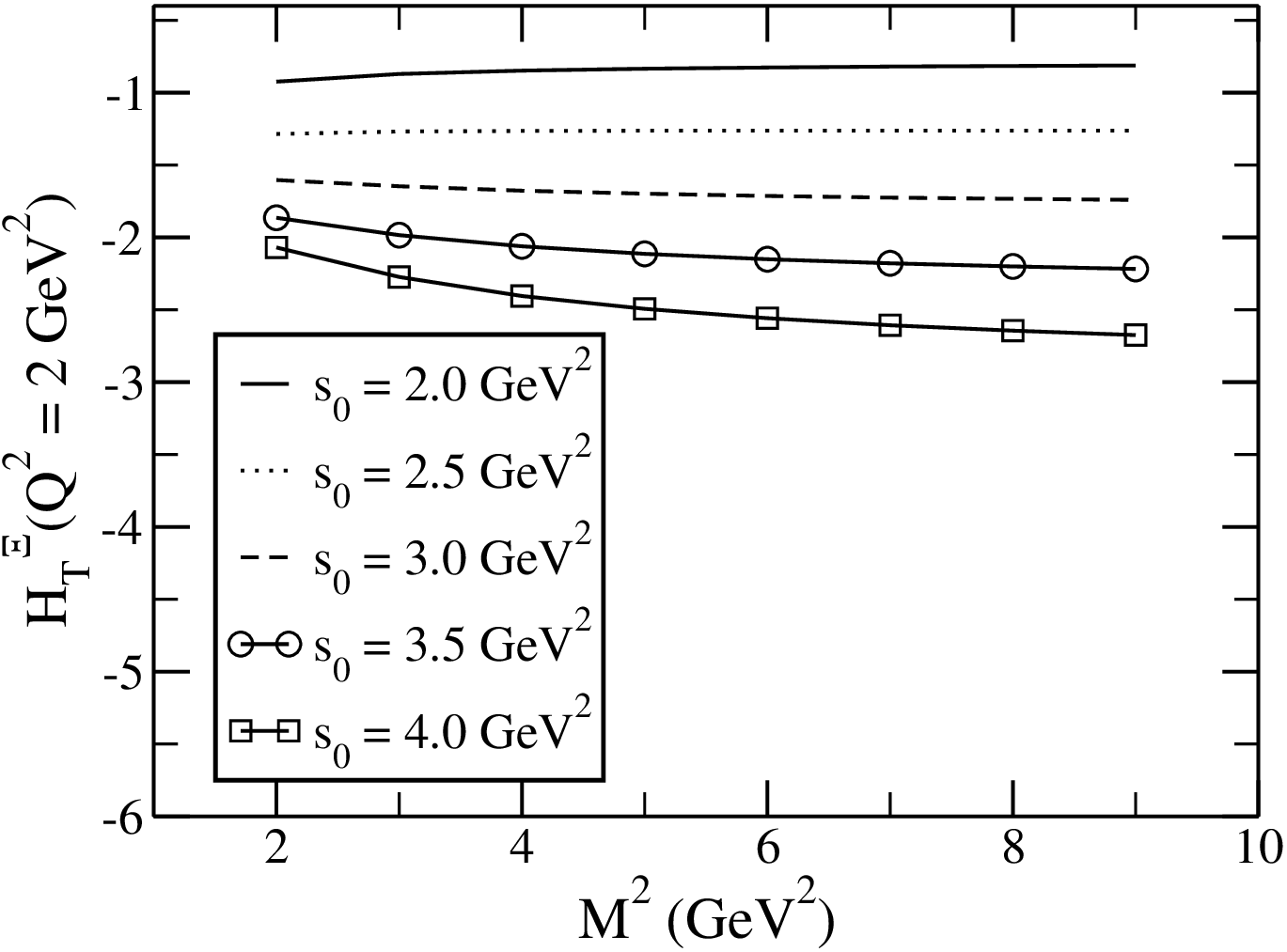}}\\
  \subfloat[]{\label{fig:ChiHtilMsq.eps}\includegraphics[width=0.35\textwidth]{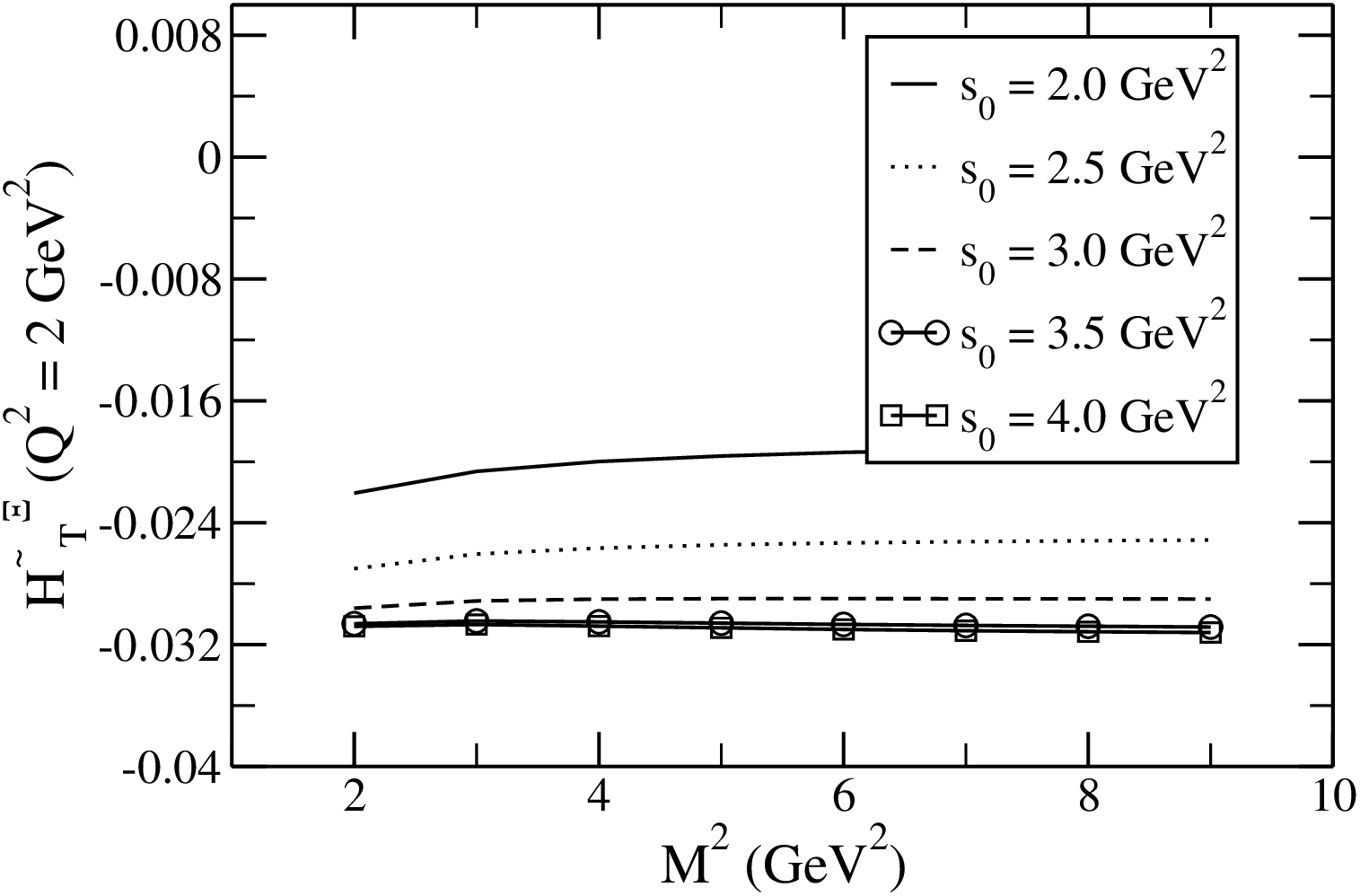}}\\
  \caption{ The dependence of the form factors; on the Borel parameter squared $M^{2}$
  for the values of the continuum threshold $s_0 = 2 ~GeV^2$, $s_0 = 2.5~GeV^2$,  $s_0 = 3 ~GeV^2$,
  $s_0 = 3.5~GeV^2$ and $s_0 = 4~GeV^2$ and $Q^2 = 2~GeV^2$,
    (a)  for $E_T^\Xi$ tensor form factor,
    (b)  for $H_T^\Xi$ tensor form factor,
    (c)  for $\tilde H_T^\Xi$ tensor form factor.}
\end{figure}
\begin{figure}[htp]
\centering
 \subfloat[]{\label{fig:SigEt.eps}\includegraphics[width=0.35\textwidth]{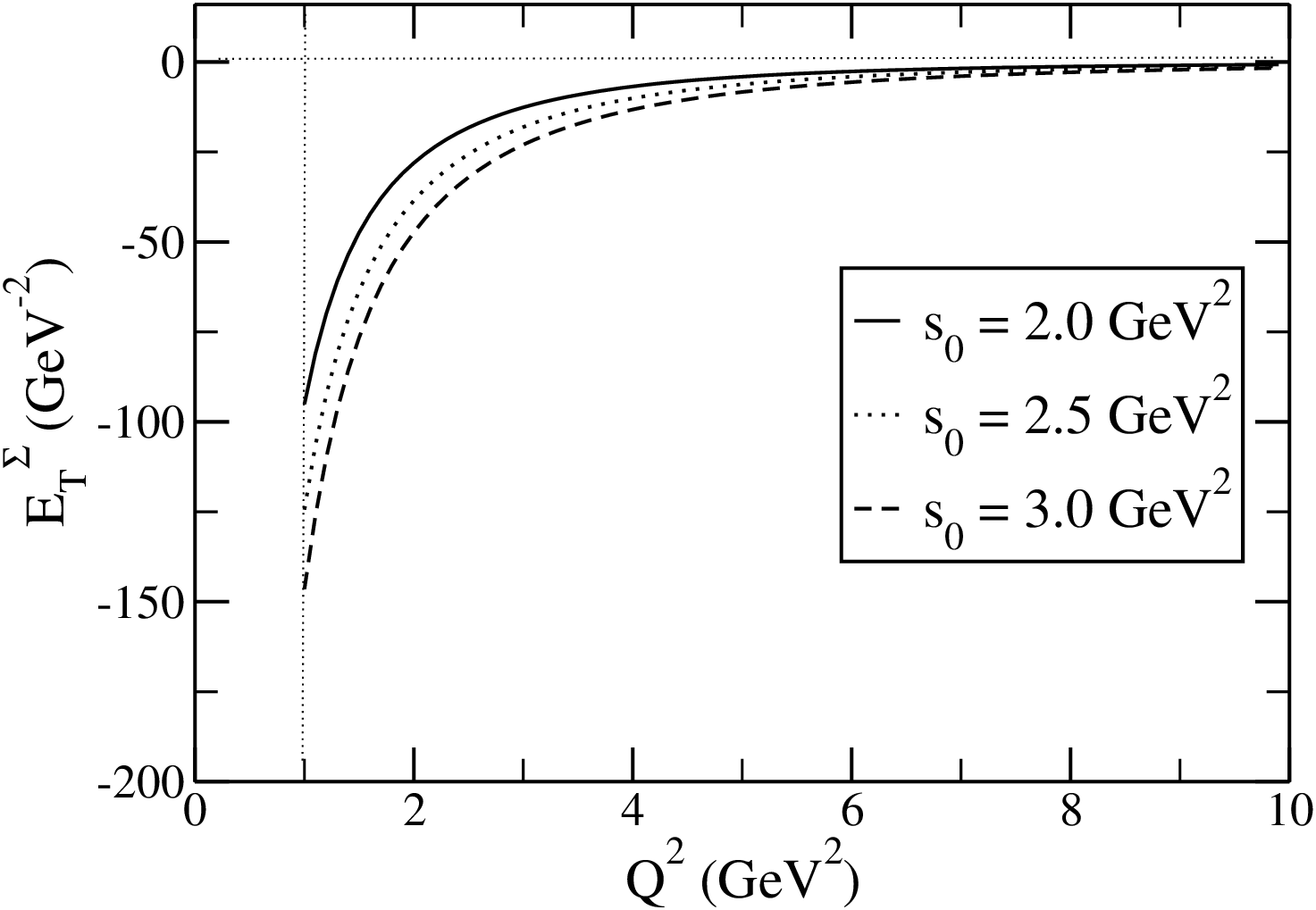}}
\subfloat[]{\label{fig:OSigEt.eps}\includegraphics[width=0.35\textwidth]{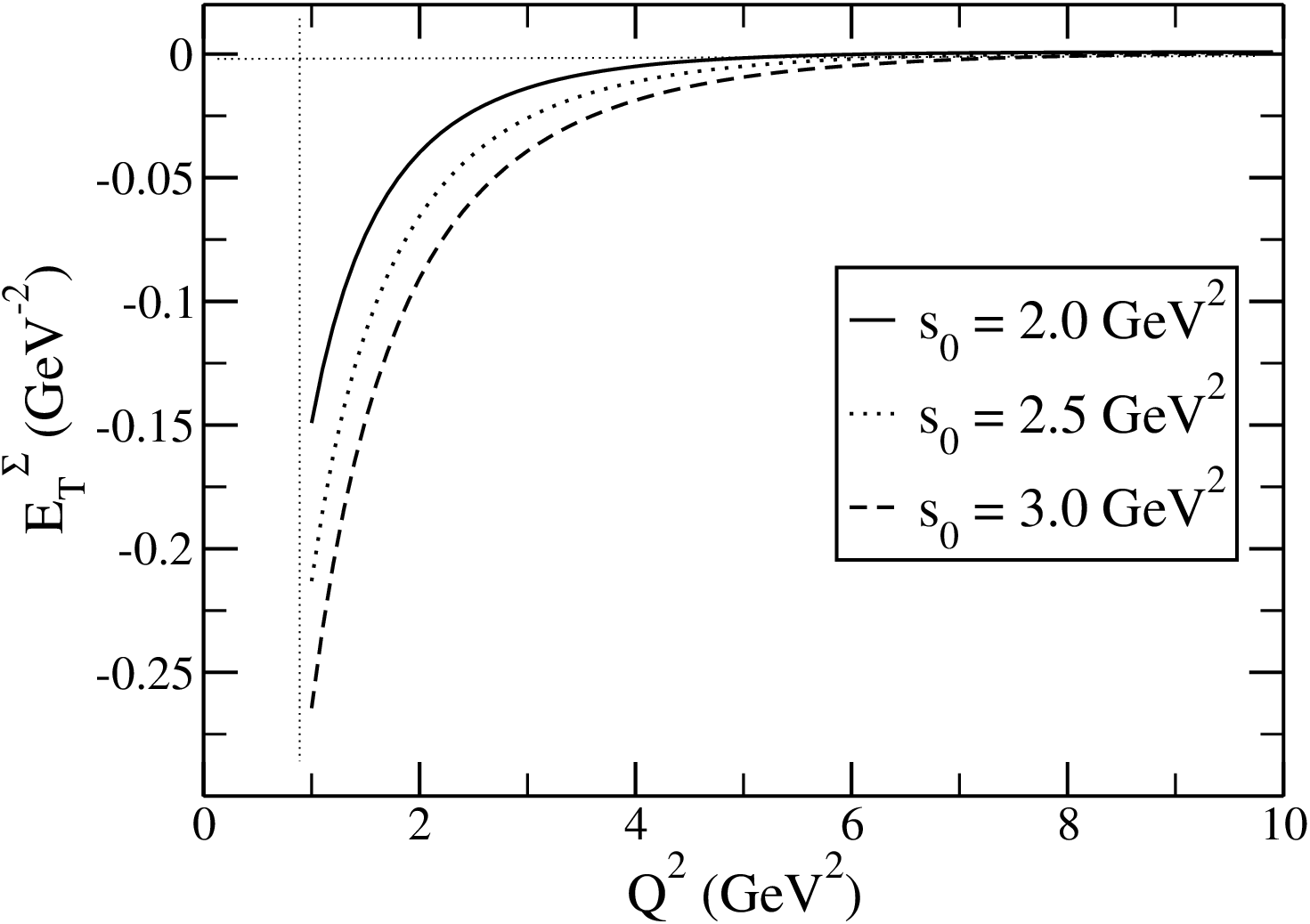}}\\ 
 \subfloat[]{\label{fig:SigHt.eps}\includegraphics[width=0.35\textwidth]{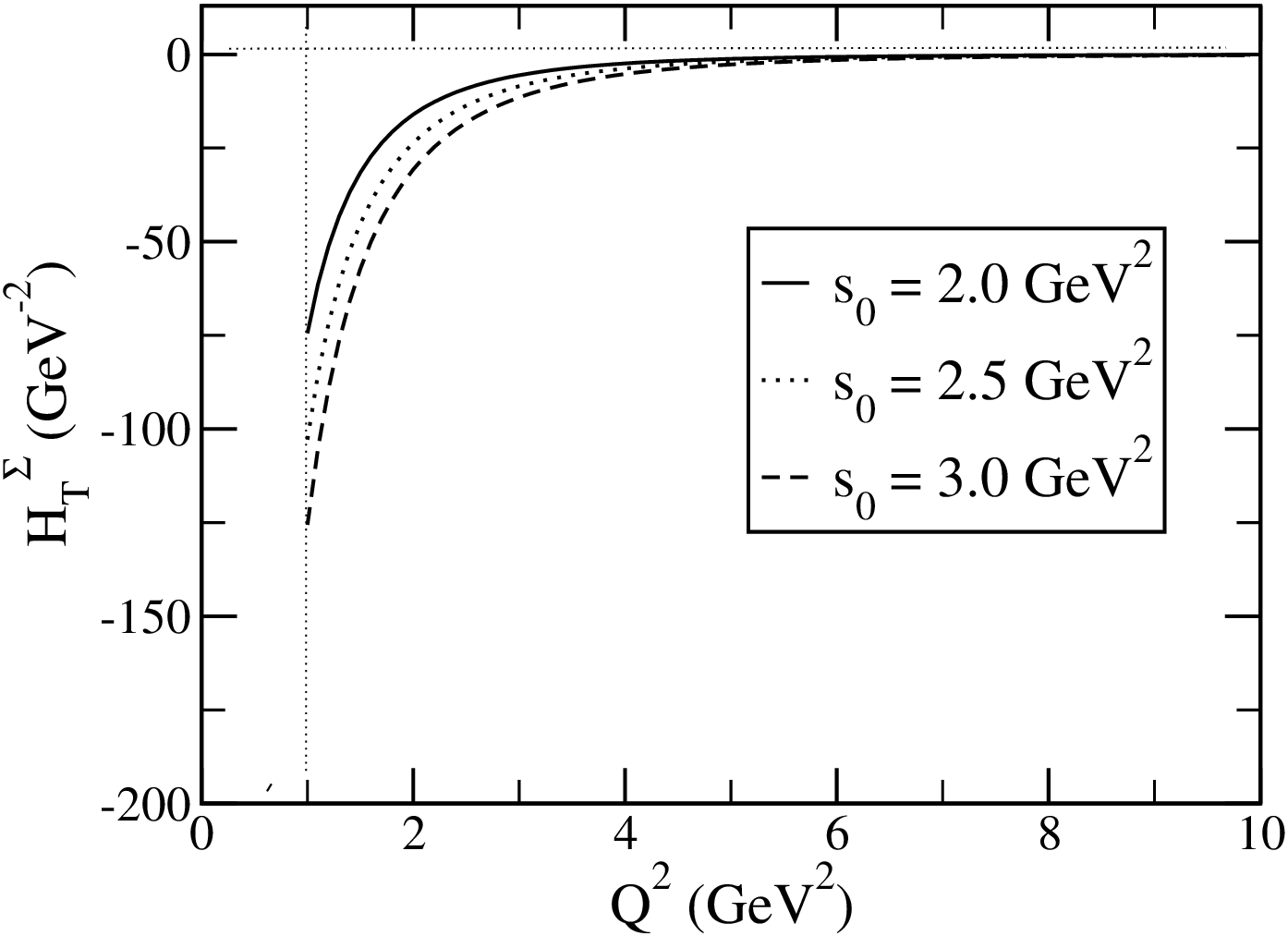}}
\subfloat[]{\label{fig:OSigHt.eps}\includegraphics[width=0.35\textwidth]{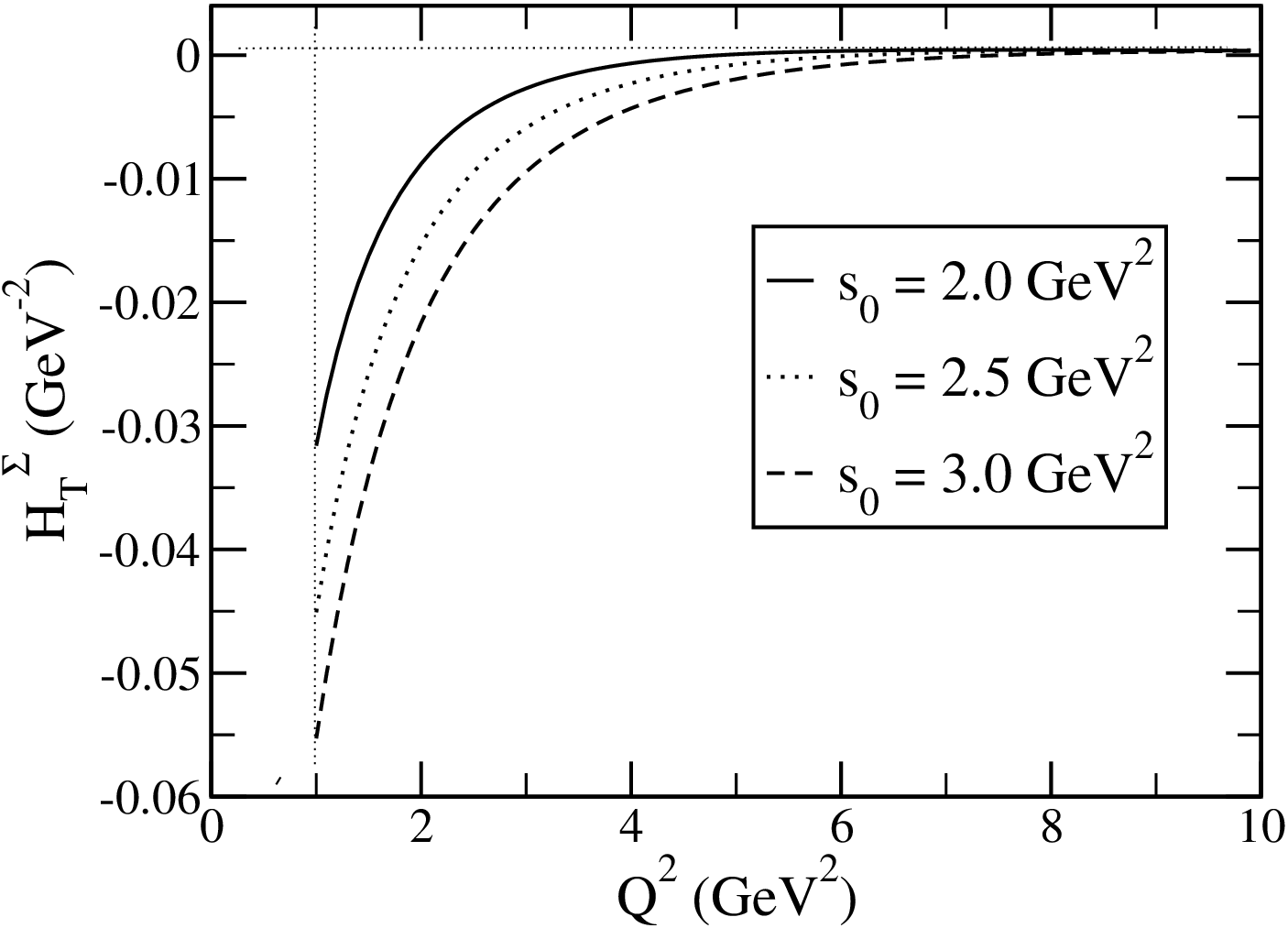}}\\
   \subfloat[]{\label{fig:SigHtil.eps}\includegraphics[width=0.35\textwidth]{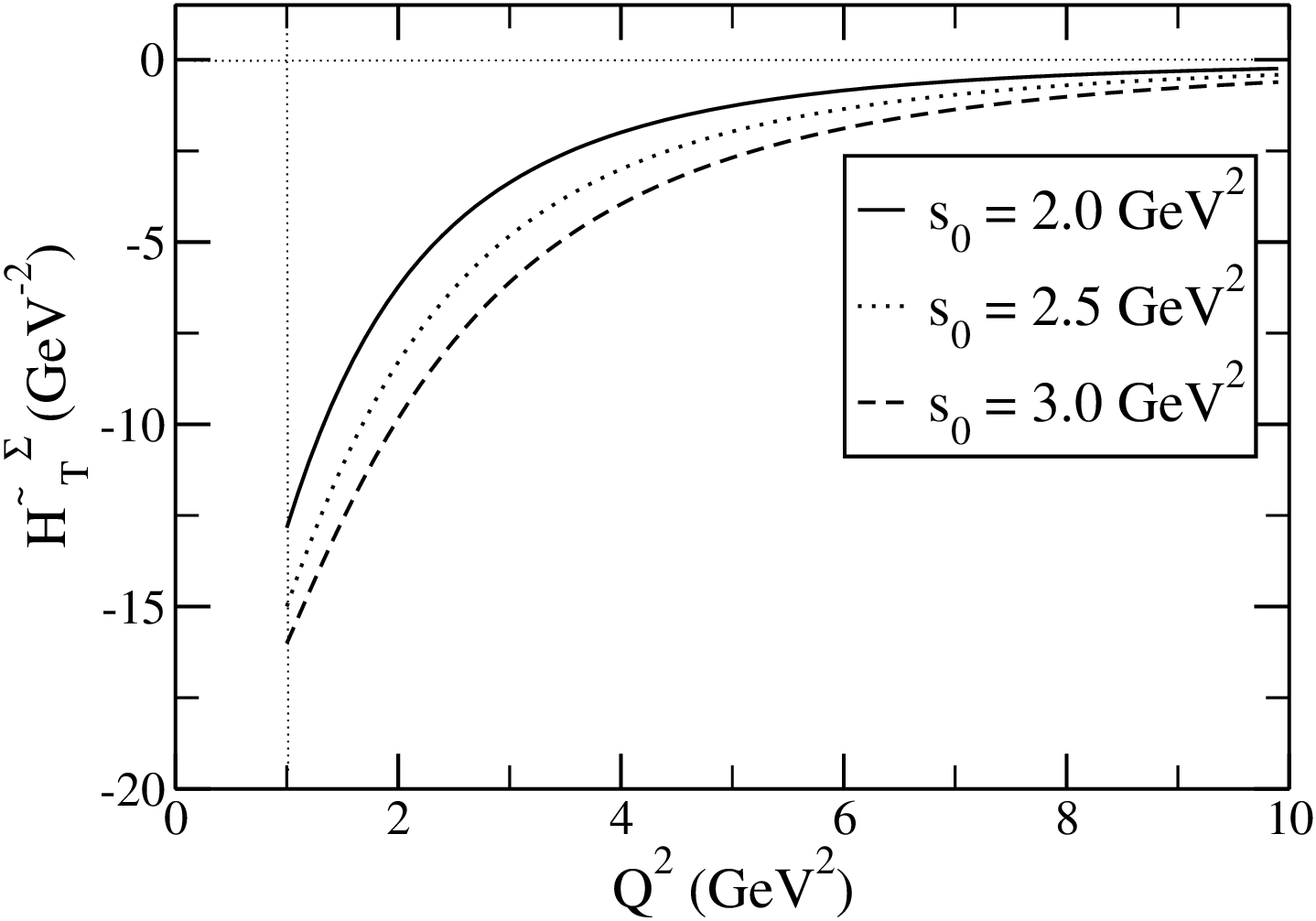}}
   \subfloat[]{\label{fig:OSigHtil.eps}\includegraphics[width=0.35\textwidth]{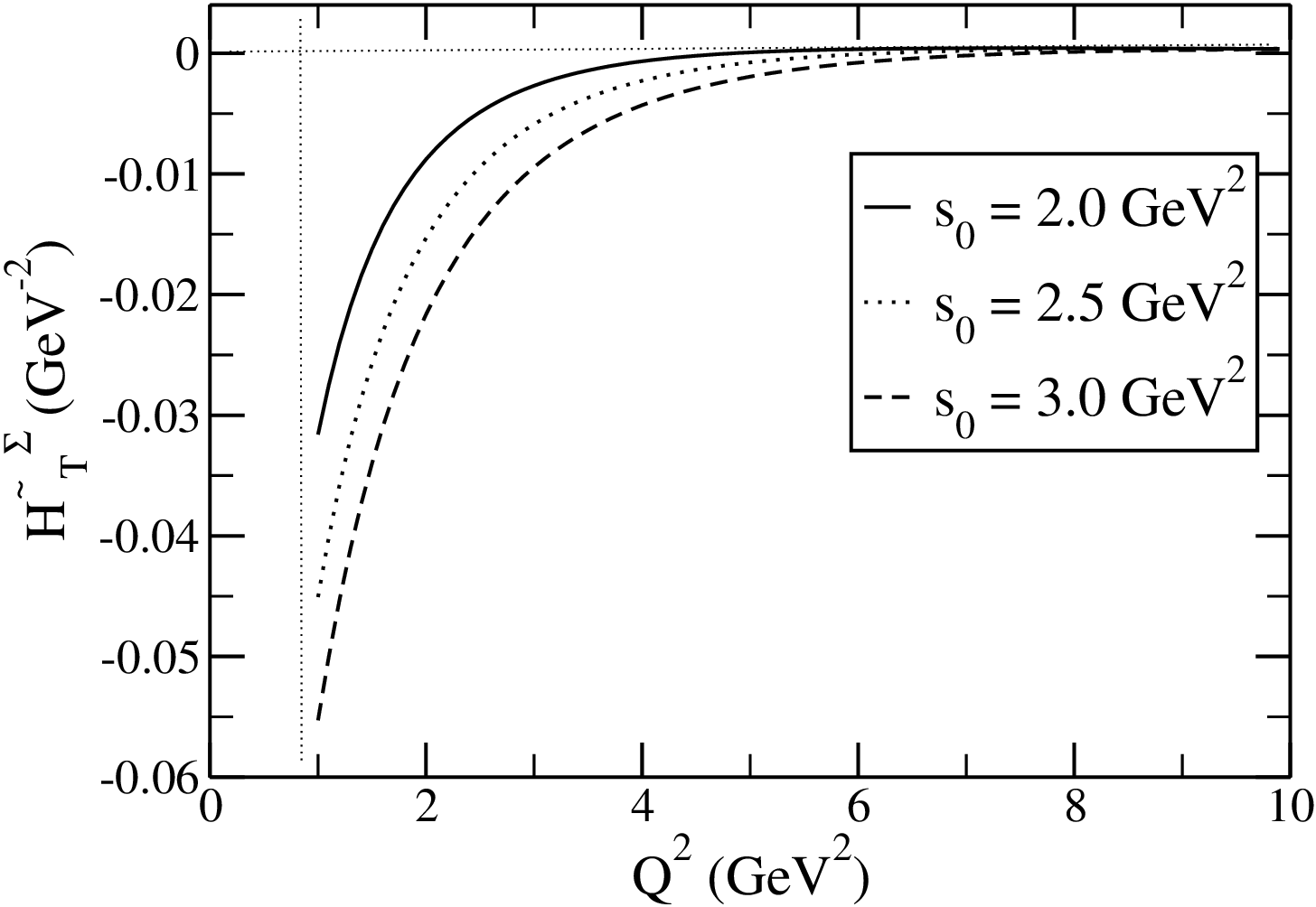}}
 \caption{The dependence of the  form factors  on the values of the continuum threshold
 $s_0 = 2 ~GeV^2$, $s_0 = 2.5~GeV^2$,  $s_0 = 3 ~GeV^2$ and $M^{2}=3~GeV^2$,
    (a) and (b) for $E_T^\Sigma$ tensor form factor,
    (c) and (d) for $H_T^\Sigma$ tensor form factor,
    (e) and (f) for $\tilde H_T^\Sigma$ tensor form factor.  In here (a), (c) and (e) are represent results of new DAs and (b), (d) and (f) are represent result of old DAs.}
\end{figure}

\begin{figure}[htp]
\centering
 \subfloat[]{\label{fig:lamSEt.eps}\includegraphics[width=0.35\textwidth]{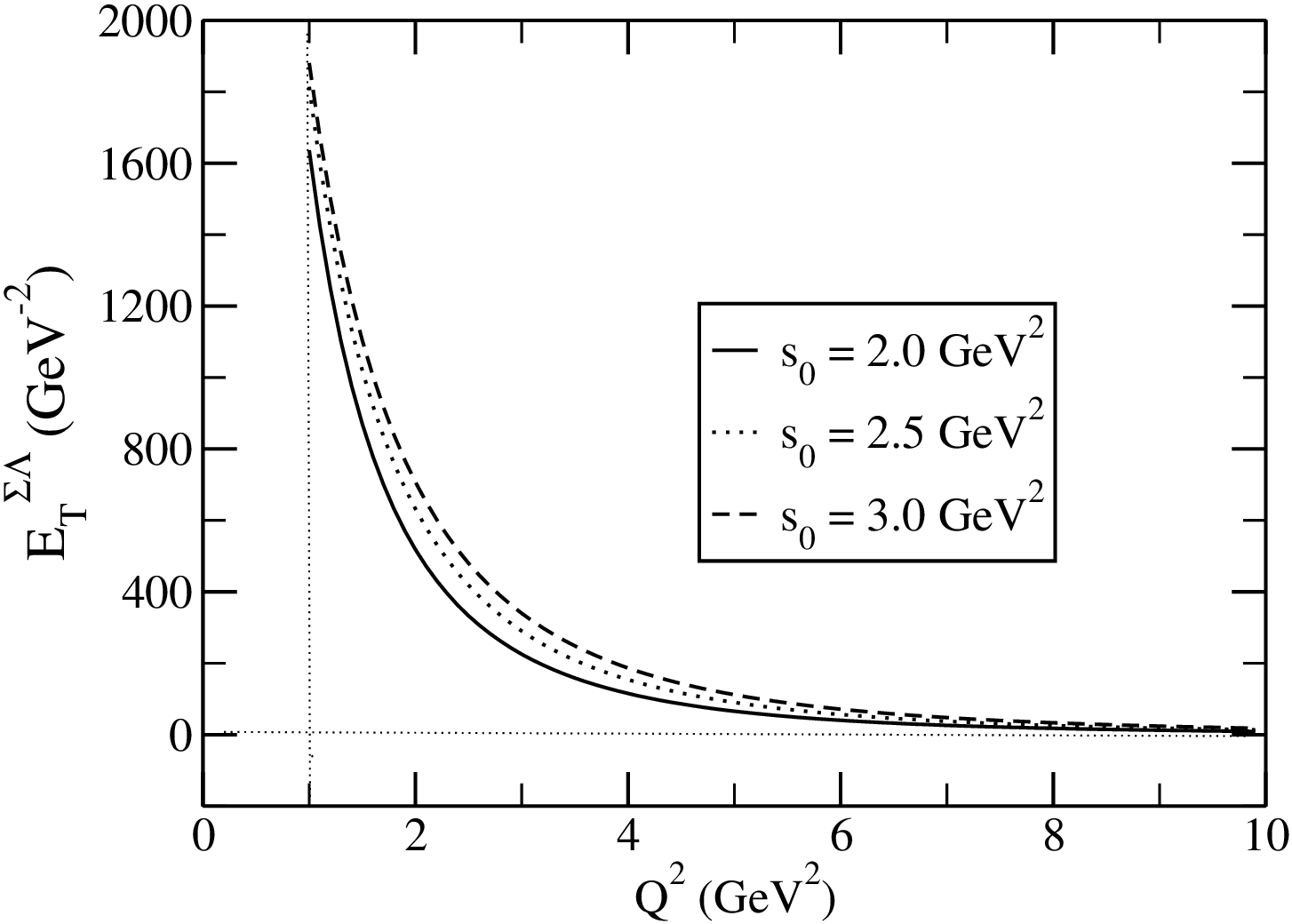}}
 \subfloat[]{\label{fig:OlamSEt.eps}\includegraphics[width=0.35\textwidth]{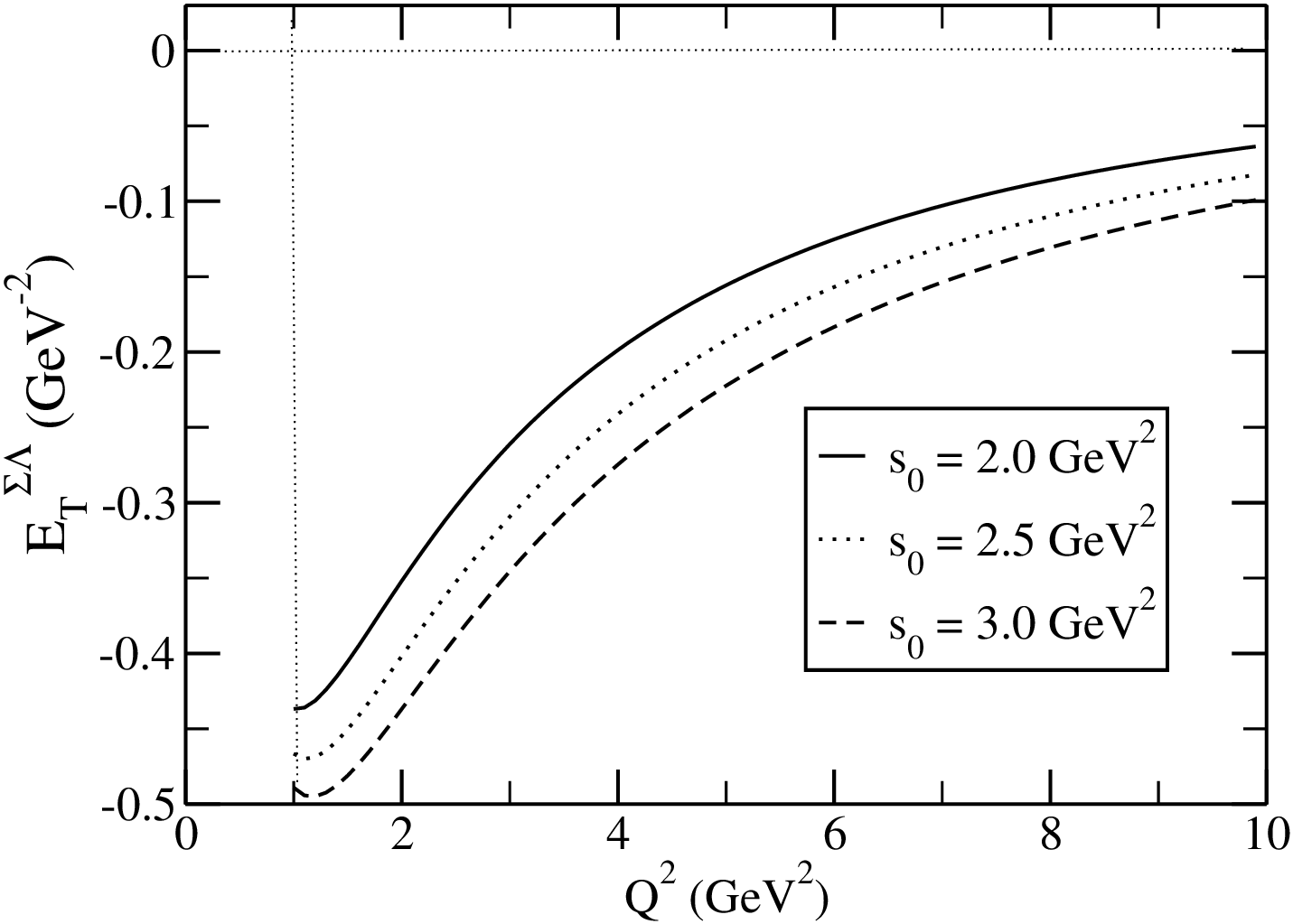}}\\
  \subfloat[]{\label{fig:lamSHt.eps}\includegraphics[width=0.35\textwidth]{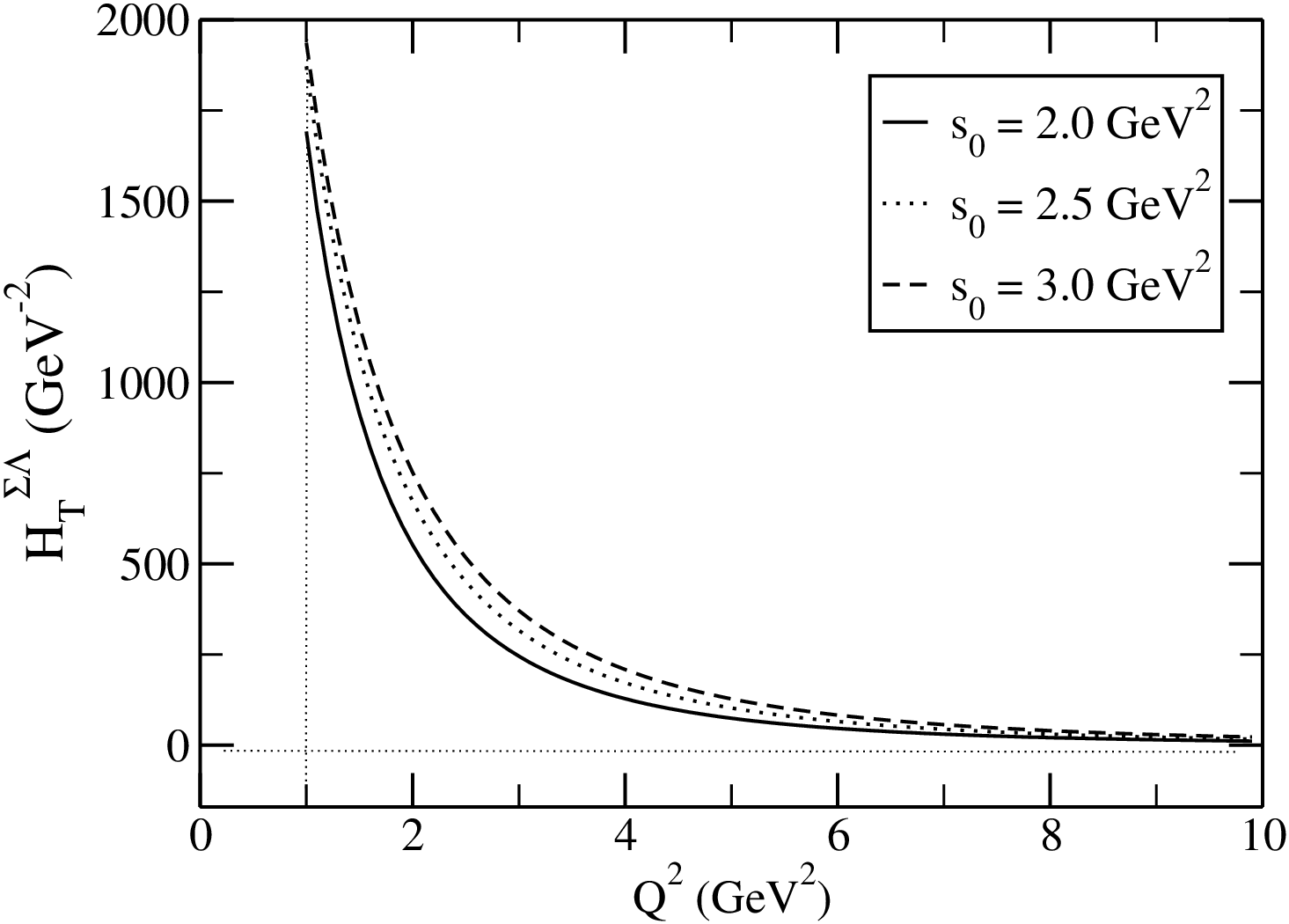}}
  \subfloat[]{\label{fig:OlamSHt.eps}\includegraphics[width=0.35\textwidth]{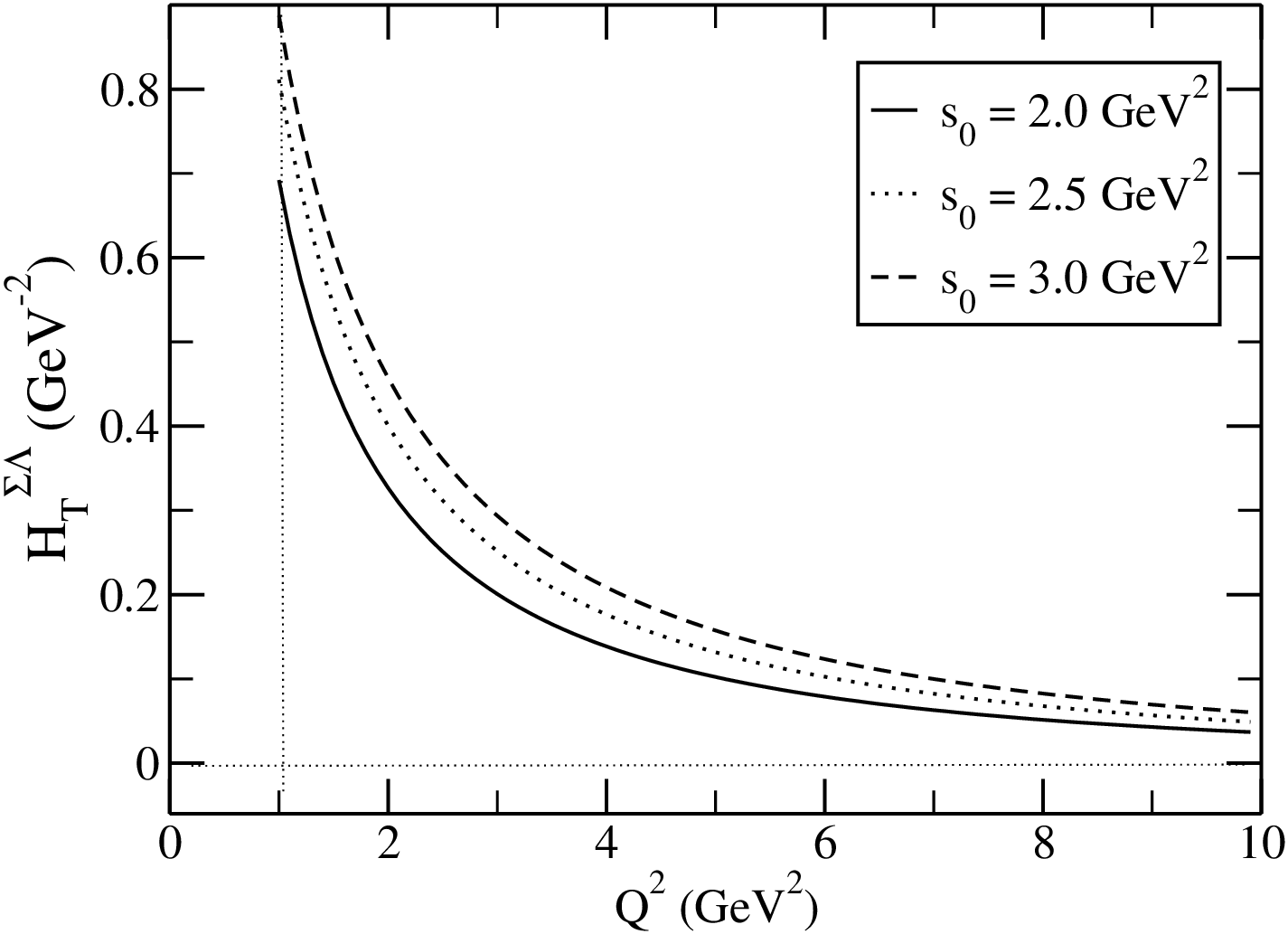}}\\
  \subfloat[]{\label{fig:lamSHtil.eps}\includegraphics[width=0.35\textwidth]{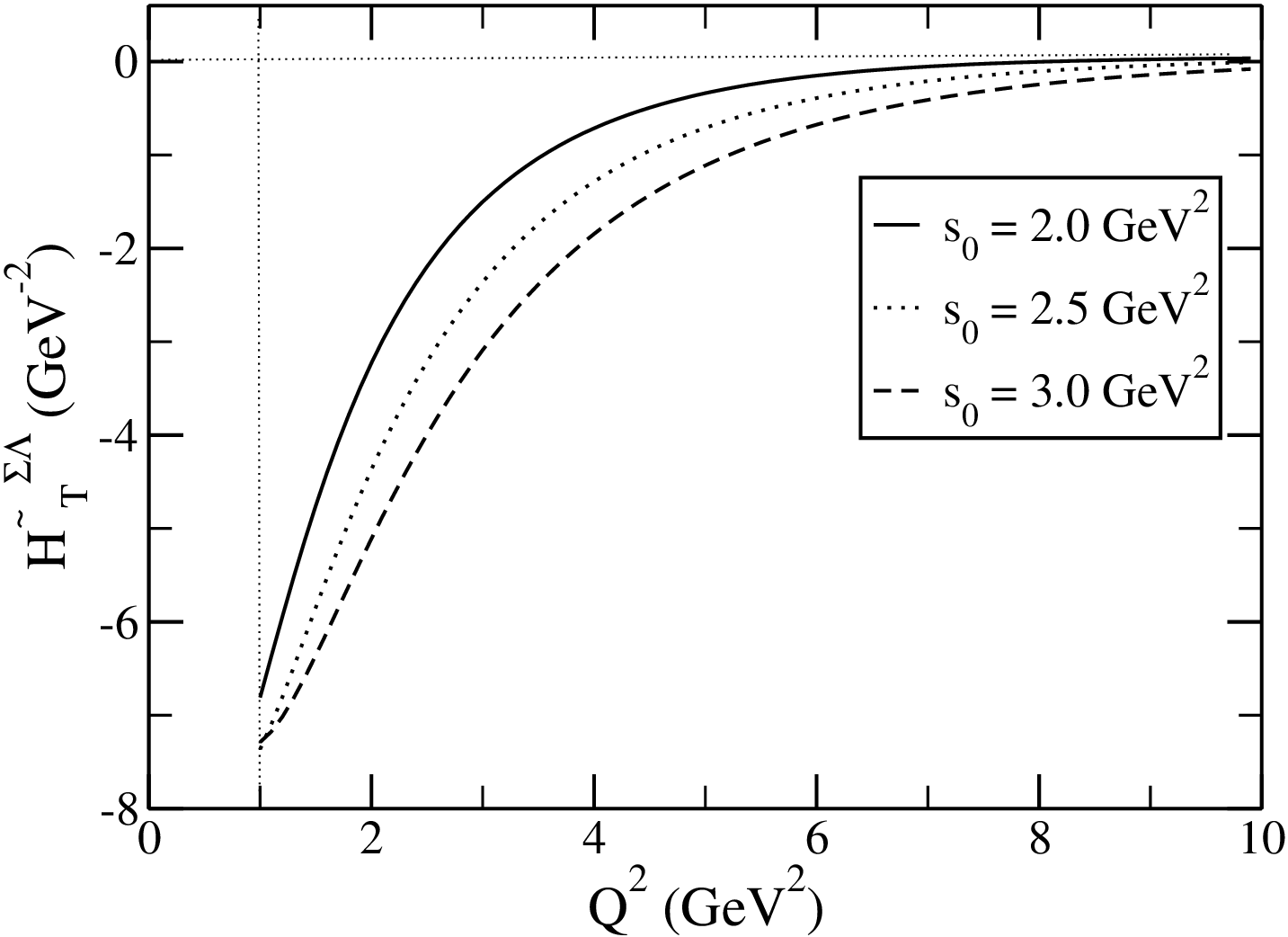}}
  \subfloat[]{\label{fig:OlamSHtil.eps}\includegraphics[width=0.35\textwidth]{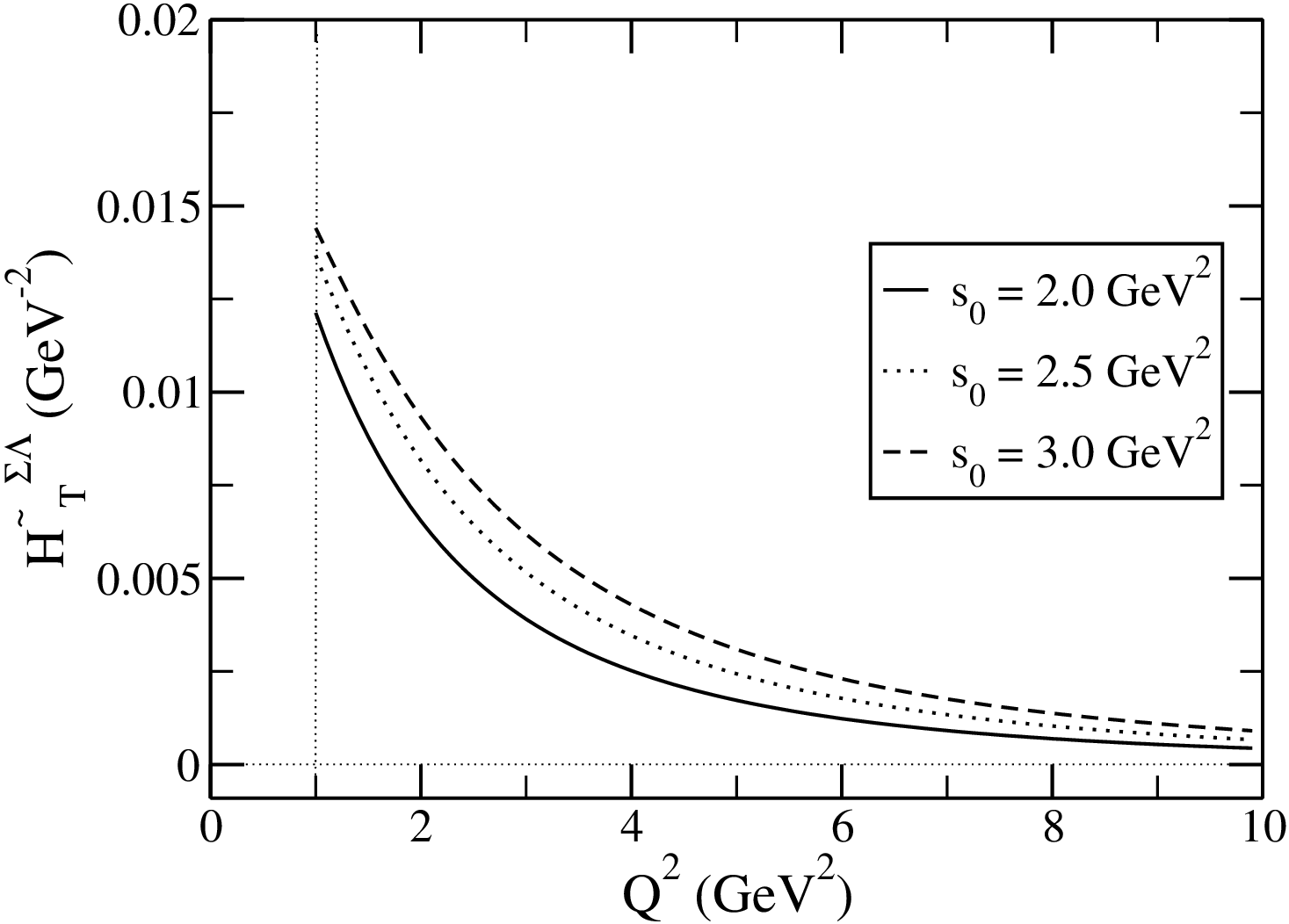}}
  \caption{The dependence of the  form factors  on the values of the continuum threshold
$s_0 = 2 ~GeV^2$, $s_0 = 2.5~GeV^2$,  $s_0 = 3 ~GeV^2$ and $M^{2}=3~GeV^2$,
    (a) and (b) for $E_T^{\Sigma \Lambda}$  form factor,
    (c) and (d) for $H_T^{\Sigma \Lambda}$ tensor form factor,
    (e) and (f) for $\tilde H_T^{\Sigma \Lambda}$ tensor form factor.  In here (a), (c) and (e) are represent results of new DAs and (b), (d) and (f) are represent result of old DAs.}
\end{figure}
\begin{figure}[htp]
\centering
 \subfloat[]{\label{fig:ChiEt.eps}\includegraphics[width=0.35\textwidth]{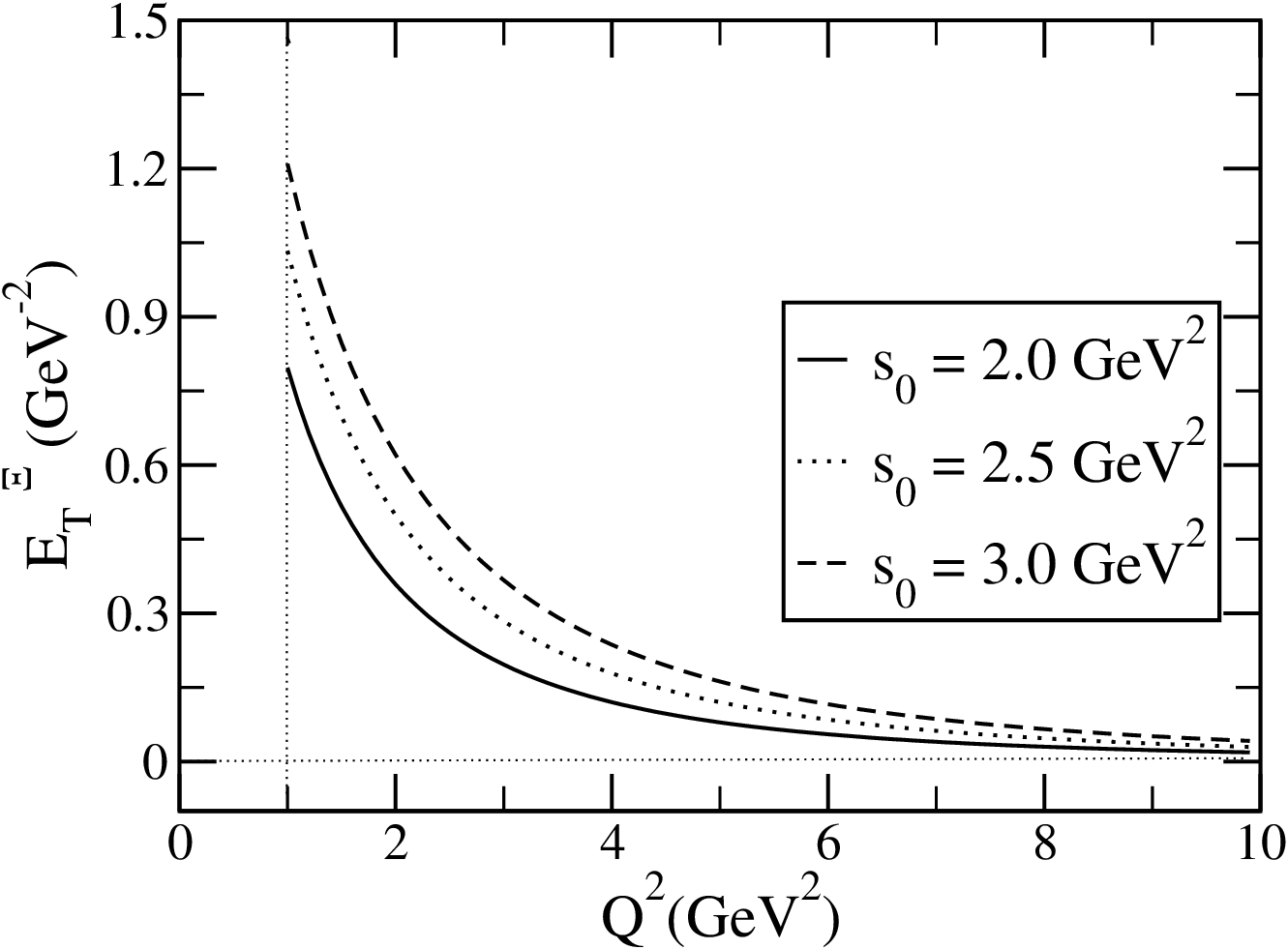}}\\
  \subfloat[]{\label{fig:ChiHt.eps}\includegraphics[width=0.35\textwidth]{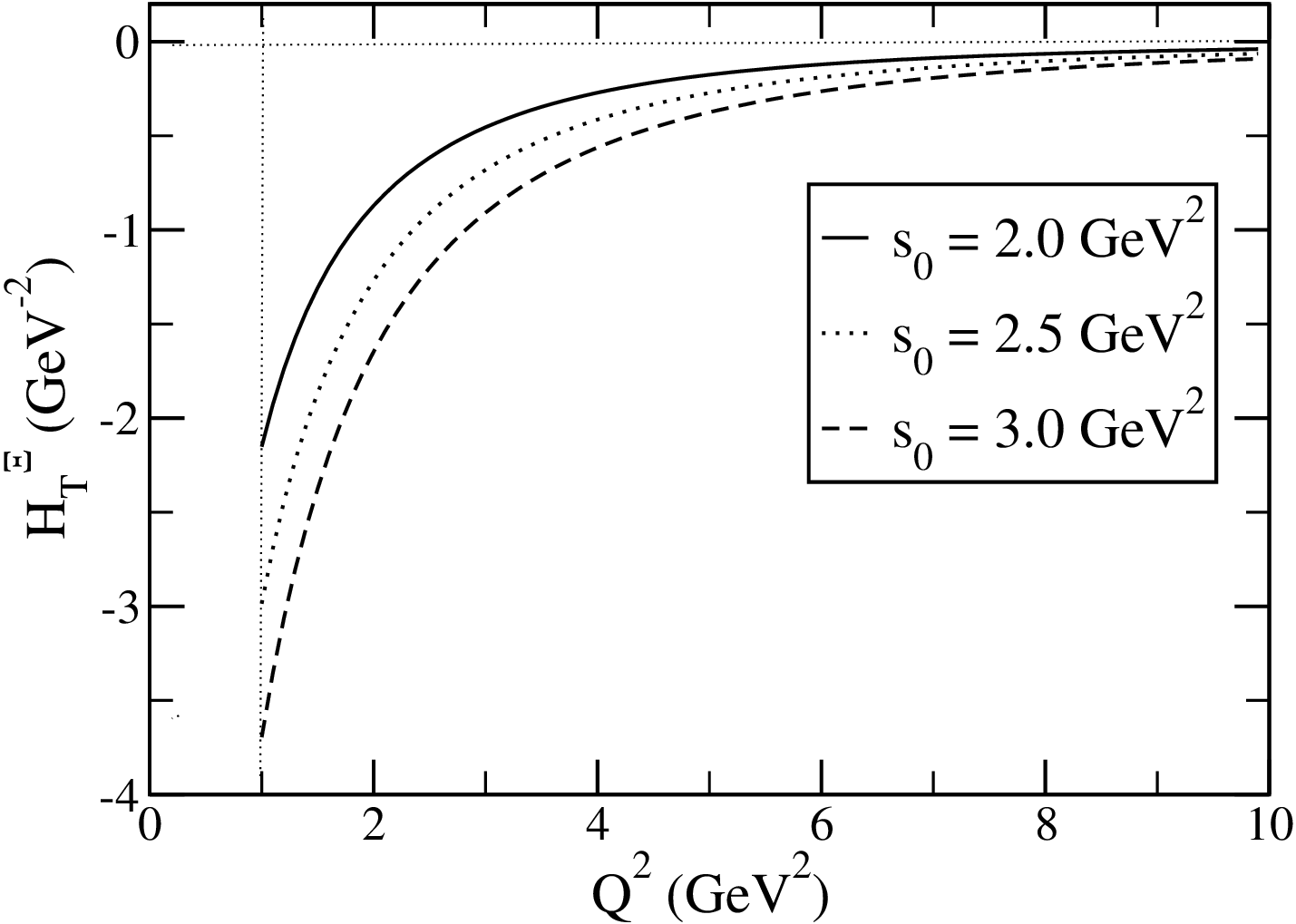}}\\
  \subfloat[]{\label{fig:ChiHtil.eps}\includegraphics[width=0.35\textwidth]{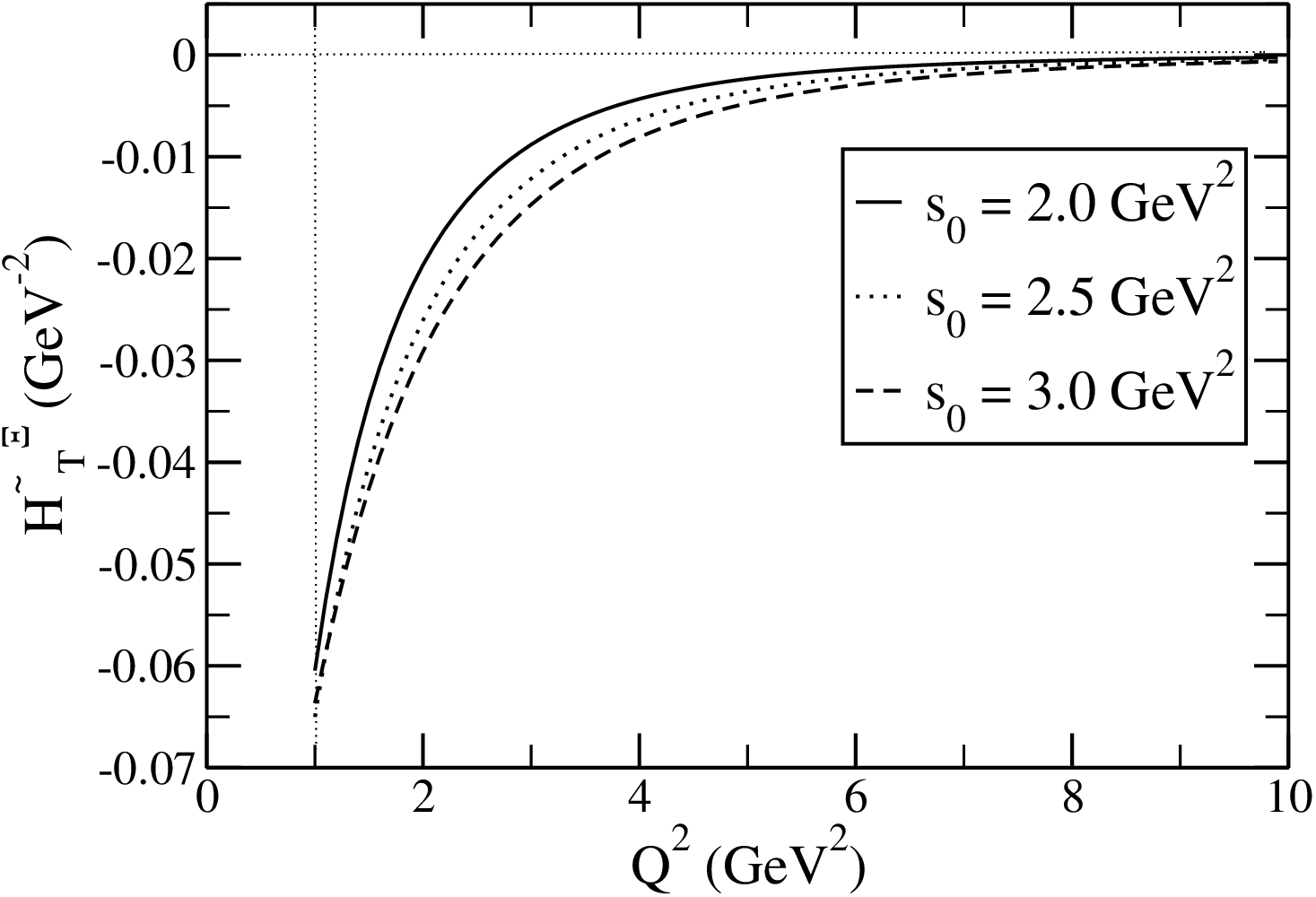}}
  \caption{The dependence of the  form factors  on the values of the continuum threshold
$s_0 = 2 ~GeV^2$, $s_0 = 2.5~GeV^2$,  $s_0 = 3 ~GeV^2$ and $M^{2}=3~GeV^2$,
    (a) for $E_T^\Xi$ tensor form factor,
    (b) for $H_T^\Xi$ tensor form factor,
    (c) for $\tilde H_T^\Xi$ tensor form factor.}
\end{figure}
\end{document}